\documentclass[a4paper]{PoS}

\usepackage{url}
\usepackage{epsfig}
\usepackage{amssymb}
\usepackage{amsmath}
\usepackage{mathrsfs}
\usepackage{multirow}
\usepackage{feynmp-auto}
\usepackage[makeroom]{cancel}

\def\be{\begin{equation}}
\def\ee{\end{equation}}
\def\ba{\begin{eqnarray}}
\def\ea{\end{eqnarray}}
\newcommand{\bea}{\begin{eqnarray}}
\newcommand{\eea}{\end{eqnarray}}

\def\Li{\textrm{Li}}

\def\eqn#1{eq.~(\ref{#1})}

\def\eqn#1{eq.~(\ref{#1})}

\newcommand{\fwboxL}[2]{\text{\makebox[#1][l]{$#2$}}}

\def\Disc{{\rm Disc}}
\def\Gcusp{\Gamma_{\rm cusp}}

\newcommand{\cA}{\begin{cal}A\end{cal}}

\newcommand{\cE}{\begin{cal}E\end{cal}}

\newcommand{\cG}{\begin{cal}G\end{cal}}
\newcommand{\cH}{\begin{cal}H\end{cal}}

\newcommand{\cN}{\begin{cal}N\end{cal}}
\newcommand{\cO}{\begin{cal}O\end{cal}}
\newcommand{\cP}{\begin{cal}P\end{cal}}

\newcommand{\cR}{\begin{cal}R\end{cal}}

\newcommand{\cZ}{\begin{cal}Z\end{cal}}

\newcommand\ns[1]{\text{ns}[#1]}
\newcommand\hns[1]{\text{hns}[#1]}

\newcommand\hnsqbar[1]{\text{hns}_{\bar Q}[#1]}

\newcommand{\la}{\langle}
\newcommand{\ra}{\rangle}
\newcommand{\ab}[1]{\la #1 \ra}

\def\green#1{{\color{green}#1}}

\usepackage{tikz}
\usepackage{pgf}
\usepackage{tcolorbox}
\usetikzlibrary{tikzmark}

\usetikzlibrary{decorations.pathmorphing}
\usetikzlibrary{intersections}
  \usetikzlibrary{positioning}
\usetikzlibrary{shapes}
\usetikzlibrary{fit}
\usetikzlibrary{backgrounds}

\usetikzlibrary{decorations.pathreplacing,calc,arrows,decorations.markings}

\title{The Steinmann Cluster Bootstrap for\\
$\mathcal{N}=4$ Super Yang-Mills Amplitudes}

\ShortTitle{Steinmann Cluster Bootstrap for $\mathcal{N}=4$ SYM Amplitudes}

\author{Simon~Caron-Huot,$^1$ Lance~J.~Dixon,$^{2}$ James~M.~Drummond,$^3$ Falko Dulat,$^{2}$ Jack~Foster,$^3$ \"Omer~G\"urdo\u gan,$^{3,4}$ Matt~von~Hippel,$^{5,6}$ Andrew~J.~McLeod$^{2,6}$ and \speaker{Georgios~Papathanasiou}\,$^{7}$\\
\llap{$^1$} Department of Physics, McGill University, 
3600 Rue University, Montr\'eal, QC Canada\\
\llap{$^2$} SLAC National Accelerator Laboratory,
Stanford University, Stanford, CA 94309, USA\\
\llap{$^3$} School of Physics \& Astronomy, University of Southampton, Southampton, SO17 1BJ, UK.\\
\llap{$^4$} Mathematical Institute, University of Oxford, Oxford, OX2 6GG UK\\
\llap{$^5$} Perimeter Institute for Theoretical Physics, 
Waterloo, Ontario N2L 2Y5, Canada\\
\llap{$^6$} Niels Bohr International Academy, Blegdamsvej 17, 2100 Copenhagen, Denmark\\
\llap{$^{7}$} DESY Theory Group, DESY Hamburg, Notkestra{\ss}e 85, D-22607 Hamburg, Germany

\vspace{4pt}

Email:\email{schuot@physics.mcgill.ca}, \email{lance@slac.stanford.edu}, \email{j.m.drummond@soton.ac.uk}, \email{dulatf@slac.stanford.edu}, \email{j.a.foster@soton.ac.uk}, \email{omer.gurdogan@maths.ox.ac.uk} \email{mvonhippel@nbi.ku.dk}, \email{amcleod@nbi.ku.dk}, \email{georgios.papathanasiou@desy.de}
}

\abstract{We review the bootstrap method for constructing six- and seven-particle amplitudes in planar $\mathcal{N}=4$ super Yang-Mills theory, by exploiting their analytic structure. We focus on  two recently discovered properties which greatly simplify this construction at symbol and function level, respectively: the extended Steinmann relations, or equivalently cluster adjacency, and the coaction principle. We then demonstrate their power in determining the six-particle amplitude through six and seven loops in the NMHV and MHV sectors respectively, as well as the symbol of the NMHV seven-particle amplitude to four loops. }

\FullConference{Corfu Summer Institute 2019 "School and Workshops on Elementary Particle Physics and Gravity" (CORFU2019)\\
		31 August - 25 September 2019\\
		Corfu, Greece}

\begin{document}

\section{Introduction}
Inspired by the analytic S-matrix bootstrap of the 60's \cite{ELOP}, the recent reincarnation of the bootstrap philosophy in perturbative quantum field theory aims at constructing physical quantities directly from the knowledge of their expected analytic structure, without ever having to resort to Feynman diagrams. It has been developed to the highest perturbative order for $n$-particle amplitudes in planar~\cite{tHooft:1973alw} $\cN=4$ Super Yang Mills (SYM) theory~\cite{Brink:1976bc,Gliozzi:1976qd} with $n=6,7$~\cite{Dixon:2011pw,Dixon:2011nj,Dixon:2013eka,Dixon:2014voa,Dixon:2014iba,Drummond:2014ffa,Dixon:2015iva,Caron-Huot:2016owq,Dixon:2016nkn,Drummond:2018caf,Caron-Huot:2019vjl,Caron-Huot:2019bsq}, following a remarkable earlier conjecture on the all-loop structure of amplitudes for any $n$ by Bern, Dixon and Smirnov~\cite{Bern:2005iz}, as well as the exploration of its consequences at strong coupling, that Alday, Maldacena and subsequent collaborators carried out \cite{Alday:2007hr,Alday:2009yn,Alday:2009dv,Alday:2010vh} via the gauge/string duality \cite{Maldacena:1997re}.

The cornerstone of the amplitude bootstrap is the construction of an ansatz for the amplitude, based on certain reasonable assumptions about the general class of functions in which it ``lives'', as well as about the specific function arguments, where the kinematic dependence enters. In its various refinements over the years, new computations go hand in hand with a deeper understanding of important properties of quantum field theory, which prune the initial space of functions containing the amplitude, thereby greatly facilitating its identification.

In the original formulation of the amplitude bootstrap at symbol \cite{Goncharov:2010jf} and polylogarithmic function ~\cite{Chen,G91b,Goncharov:1998kja} level, it was the property of physical singularities, as dictated by unitarity~\cite{Gaiotto:2011dt}, that was key for obtaining the six-particle maximally helicity violating (MHV) amplitude at 3 and 4 loops \cite{Dixon:2011nj,Dixon:2013eka,Dixon:2014voa}, and of the next-to-maximallly helicity ivolating (NMHV) amplitude at two loops~\cite{Dixon:2011pw}. Then, it was the appreciation of the consequences of dual superconformal symmetry~\cite{Drummond:2008vq}, in the form of the $\bar Q$-equation \cite{CaronHuot:2011kk}, which led to the determination of the aforementioned NMHV amplitude at 3 and 4 loops \cite{Dixon:2014iba,Drummond:2014ffa}. In upgrading the method from six to seven particles, crucial predictions for the set of potential singularities of the amplitude came from cluster algebras \cite{Golden:2013xva}, enabling in turn the calculation of the 3-loop MHV symbol \cite{Drummond:2014ffa}. And further restrictions on the double discontinuity structure of amplitudes, stemming from the Steinmann relations \cite{Steinmann,Steinmann2,Cahill:1973qp}, were instrumental for pushing the computations to 5 loops for the six-particle amplitude~\cite{Caron-Huot:2016owq}, and to 3 and 4 loops for the NMHV and MHV helicity configurations of the seven-particle amplitude~\cite{Dixon:2016nkn}, respectively.

In this contribution, we will mainly focus on two recently discovered analytic properties, and their implications. First, we will discuss the \emph{extended Steinmann relations}, which essentially generalize the ordinary Steinmann relations, so as to hold not only for double, for also for any higher multiple discontinuity \cite{Caron-Huot:2019bsq}. Very interestingly, cluster algebras provide equivalent restrictions on multiple discontinuities, by means of the \emph{cluster adjacency} property \cite{Drummond:2017ssj,Drummond:2018dfd}. The fact that extended Steinmann/cluster adjacency holds not only for amplitudes, but also for individual Feynman integrals \cite{Caron-Huot:2018dsv}, thus opens the exciting possibility, that the amplitude bootstrap reveals, for the first time, a universal property of quantum field theory, with far wider applicability.

The second new property we will discuss is the (cosmic Galois) coaction principle~\cite{Cartier2001,Schnetz:2013hqa,Brown:2015fyf,Panzer:2016snt}, which is based on an established operation for decomposing elements in the space of functions containing the amplitude into simpler building blocks, known as the coaction \cite{Gonch3,Gonch2,Brown:2011ik,Duhr:2012fh}. The coaction principle then asserts that, loosely speaking, an element in this space can be a candidate amplitude or a derivative thereof, only if its building blocks have already appeared as amplitudes or their derivatives at lower loops. In this manner, it provides significant guidance towards constructing the \emph{minimal} relevant function space that only contains amplitudes and their derivatives, and thus maximally simplifies their determination.

Having presented the extended Steinmann/cluster adjacency and coaction principle properties, we will then move on to present new results based on their application: These are the 6-loop NMHV and 7-loop MHV six-particle amplitude, as well as the symbol of the 4-loop NMHV seven-particle amplitude. This contribution to the proceedings of CORFU2019 is based on the publications~\cite{Drummond:2018caf,Caron-Huot:2019vjl,Caron-Huot:2019bsq}, where many more details and background material may be found. For an earlier review of the amplitude bootstrap see \cite{Dixon:2014xca}, as well as the review sections of \cite{Dixon:2016nkn,Drummond:2018dfd}.

Before concluding this introduction, let us also briefly mention alternative approaches for the computation of planar $\cN=4$ SYM amplitudes. The first nontrivial amplitude at six points and two loops was obtained analytically by means of Feynman diagrams in simplified quasi-multi-Regge kinematics \cite{DelDuca:2009au,DelDuca:2010zg}, which nevertheless provides the full answer in general kinematics due to the dual conformal symmetry of the theory \cite{Drummond:2006rz,Alday:2007hr,Drummond:2007aua,Drummond:2007au}. Other formulations of the integrand of the theory are also amenable to direct integration \cite{Bourjaily:2019vby}. In addition, apart from bootstrap ``boundary data'', the $\bar Q$-equation~\cite{CaronHuot:2011kk} additionally provides alternative representations for amplitudes, as integrals over a collinear limit of amplitudes with higher multiplicity and MHV degree, and lower loop order. These representations have been successfully evaluated at the level of the symbol, for all 2-loop MHV amplitudes \cite{CaronHuot:2011ky} (which was also promoted to a function for $n=7$ in \cite{Golden:2014xqf}), as well as the 2-loop NMHV 8-particle amplitude \cite{Zhang:2019vnm}, more recently.

Finally, the integrability-based Pentagon Operator Product Expansion (OPE)~\cite{Alday:2010ku,Basso:2013vsa,Basso:2013aha,Papathanasiou:2013uoa,Basso:2014koa,Papathanasiou:2014yva,Basso:2014nra,Belitsky:2014sla,Belitsky:2014lta,Basso:2014hfa,Basso:2015rta,Basso:2015uxa,Belitsky:2016vyq} is tantamount to an infinite kinematic expansion around the collinear limit, where every term in the expansion has a known Mellin-Barnes-like integral representation. At multiplicity $n=6$, the integrand is completely known, and thus it is in principle possible to resum this kinematic expansion, at least order by order in perturbation theory. This resummation was initiated in \cite{Drummond:2015jea}, see also \cite{Cordova:2016woh,Lam:2016rel,Bork:2019aud}. At higher multiplicity, this is no longer possible in general (but see \cite{Belitsky:2017wdo,Belitsky:2017pbb} for some tree-level exceptions), due to the fact that the so-called matrix part of the integrand is only known in recursive, and not closed, form \cite{Belitsky:2016vyq}.

\section{Planar Amplitudes: Symmetries and Kinematics}\label{sec:SymKin}
We will be focusing on 't Hooft's planar limit~\cite{tHooft:1973alw}, where the number of colors $N\to \infty$ with its product $g_{YM}^{2}N$ with the gauge theory coupling  held fixed. The latter is the only surviving parameter of the theory, and we will be denoting the perturbative expansion of any quantity $F$ with respect to it as
\be\label{eq:gLoopExpansion}
F=\sum_{L=0}^\infty g^{2L} F^{(L)}\,,\quad g^2 = \frac{{g_{YM}^{2}N}}{16 \pi^{2}}\,.
\ee
In the planar limit, a single color structure becomes dominant for $n$-gluon amplitudes, an overall trace of the gauge group generators (see for example the review~\cite{Dixon:2011xs}). We will thus be exclusively considering its coefficient, the \emph{color-ordered amplitude}. Due to its relation to the trace, it possesses a discrete \emph{dihedral symmetry}, namely it is invariant under cyclic shifts $i\to i+1$ of the particle labels, as well as under reflections $i\to n+1-i$, where the equivalence $n+i\sim i$ is understood.

We will also be restricting our attention to $\cN=4$ SYM theory~\cite{Brink:1976bc,Gliozzi:1976qd}, where all fields are massless and in the adjoint representation. Their remaining quantum numbers are therefore momenta and helicities, and thanks to the supersymmetry of the theory, the entire on-shell content of the theory can be combined in the superfield~\cite{Nair:1988bq},
\be\label{eq:PhiSuperfield}
\Phi=G^+ +\eta^A\Gamma_A+\tfrac{1}{2!}\eta^A\eta^B S_{AB}+\tfrac{1}{3!}\eta^A\eta^B\eta^C\epsilon_{ABCD}\bar \Gamma^D+\tfrac{1}{4!}\eta^A\eta^B\eta^C\eta^D\epsilon_{ABCD}G^-\,.
\ee
In the above, $G^\pm$, $\Gamma_A/\bar \Gamma^A$ and $S_{AB}$ are on-shell gluons, fermions and scalars with helicity $\pm 1,\pm 1/2$ and 0 respectively, $\eta^A$ is a Gra\ss mann variable of helicity 1/2, $A$ is a fundamental index of the $SU(4)$ R-symmetry of the theory, and $\epsilon$ is the Levi-Civita tensor, see also ~\cite{Drummond:2010km} for a review.

With the help of the superfield \eqref{eq:PhiSuperfield}, we can thus combine different amplitudes related by supersymmetry, to a single \emph{color-ordered superamplitude} $A(\Phi_1,\ldots \Phi_n)$, which also inherits the discrete symmetries of its gluon part. It will also be convenient to split this superamplitude into sectors of different \emph{helicity degree} $k$
\be
A(\Phi_1,\ldots, \Phi_n)= A_{n,0}+A_{n,1}+\ldots+A_{n,n-4}\,,\quad A_{n,k}: \text{Gra\ss mann-$\eta$ degree}\,\, 4k+8\,,
\ee
by expanding $\Phi_i$ and collecting all terms of the same power in $\eta$. From \eqref{eq:PhiSuperfield}, it is evident that $A_{n,0}$ will contain the Maximally Helicity Violating (MHV) gluon amplitudes where all but two helicities are positive. For $k<0$ or $k>n-4$ the gluon component and the entire superamplitude $A_{n,k}$ will vanish, and R-symmetry invariance also restricts the Gra\ss mann degree to be a multiple of 4.

It is also worth noting that different helicity degrees $k$ and $\bar k=n-4-k$ are related by parity or spatial reflection tranformations, which essentially amount to complex conjugation of spinors. As a result, we may always restrict $k\le  \lfloor \frac{n-4}{2}{\rfloor}$, where $\lfloor x {\rfloor}$ denotes the integer part of $x$. Thus for the cases $n=6,7$ that we will consider in this contribution, only $A_{n,0},\, A_{n,1}$, or in other words only the MHV and Next-to-MHV (NMHV) superamplitudes, will be relevant for our discussion. Evidently, amplitudes obeying $k=(n-4)/2$, and thus also $A_{6,1}$, will be invariant under parity.

Finally, let us turn to the kinematic dependence of the amplitude. The dual conformal symmetry~\cite{Drummond:2006rz,Alday:2007hr,Drummond:2007aua,Drummond:2007au} of planar $\cN=4$ SYM, which acts on dual position variables $x_i$ related to the usual momenta by
\be
p_i \equiv x_{i+1} - x_i\,,
\end{equation}
implies that amplitudes depend on fewer kinematic variables than in a generic gauge theory. In particular, after factoring out a universal infrared-divergent factor known as the BDS-like ansatz (we will come back to this point later)~\cite{Alday:2009dv}, see also~\cite{Yang:2010as}, $A_{n,k}$ depends on $3n-15$ (instead of $3n-10$) kinematic variables.\footnote{The number of kinematic variables is equal to the independent components of the points $x_i$, minus the dimension of the 4D conformal group or Lorentz group, for planar $\cN=4$ SYM or a generic massless gauge theory, respectively.} In other words, instead of the algebraically independent subset of the $n(n-3)/2$ distinct Mandelstam invariants 
\begin{equation}\label{eq:Mandelstamtox}
s_{i,\ldots,j-1}\equiv (p_{i} + p_{i+1} + \ldots + p_{j-1})^2 = (x_i - x_j)^2\equiv x^2_{ij}
\end{equation}
one must instead pick the algebraically independent subset of their distinct 
$n(n-5)/2$ \emph{conformal cross ratios}
\begin{equation}\label{u_def}
u_{ij}\equiv\frac{x_{ij+1}^2 x_{i+1j}^2}{x_{ij}^2 x_{i+1j+1}^2}\,.
\end{equation}
Particularly for $n=6,7$, the notation
\be\label{eq:udef}
u_i\equiv {u_{i-1,i+2}}
\ee
is also used. Expressing all the cross ratios in terms of an independent subset is possible with the help of (conformal) Gram determinant constraints, which simply encode the fact that no more than $d$ vectors can be linearly independent in $d$ dimensions. Nevertheless, in practice these parametrizations of the kinematics turn out to be quite complicated.

Instead, it proves significantly more advantageous to parametrize massless, planar, dual conformal invariant kinematics in terms of \emph{momentum twistors} \cite{Hodges:2009hk}, which are also reviewed for example in \cite{ArkaniHamed:2010gh}. In a nutshell, these variables stem from the fact that we can represent $x^\mu \in \mathbb{R}^{1,3}$ as projective null vectors  $X^M\in \mathbb{R}^{2,4}$, $X^2=0$, $X\sim \lambda X$. This $SO(2,4)$ vector $X^M$ is also equivalent to an antisymmetric representation $X^{IJ}$ of $SU(2,2)$, given that the two algebras are isomorphic. The latter can be built out of two copies of the fundamental representation $Z^{I}$ of $SU(2,2)$, or, after complexifying, $SL(4,\mathbb{C})$. The momentum twistors are precisely these $Z$'s, and we see that each point in $\mathbb{R}^{1,3}$ is mapped to a pair of points, namely a line in momentum twistor space. Similarly to $X$, they are also defined up to rescalings $Z\sim t Z$, and so they may be equivalently viewed as homogeneous coordinates on complex projective space $\mathbb{P}^3$. One can show that the usual Mandelstam invariants \eqref{eq:Mandelstamtox} can be expressed in terms of momentum twistors as
\begin{equation}
x^2_{ij} \propto \langle i-1 i j-1 j \rangle\,,
\end{equation}
up to proportionality factors that drop out from conformally invariant quantities, where
\be\label{fourbrak}
\ab{ijkl} \equiv\langle Z_{i} Z_j Z_{k} Z_l \rangle=\det(Z_i Z_j Z_k Z_l)
\ee
is a \emph{four-bracket} of momentum twistors. 

Conformal transformations of the dual positions $x$ map to $SO(2,4)$ rotations of $X$, and in turn to $SL(4,\mathbb{C})$ transformations of the momentum twistors. Therefore the space of dual conformal invariant kinematics can be written as a $4\times n$ matrix, whose columns are the cyclically ordered momentum twistors/homogeneous $\mathbb{CP}^3$ coordinates defined up to rescalings, and modulo $SL(4,\mathbb{C})$ transformations. In the Appendix, we provide examples of explicit parametrizations, or coordinate frames, on the space of six- and seven-particle kinematics. It is worth  noting that this space of momentum twistor kinematics is also equivalent to the quotient $Gr(4,n)/(\mathbb{C}^{*})^{n-1}$ of a Gra\ss mannian \cite{Golden:2013xva}. This may be seen by recalling that the Gra\ss mannian $Gr(m,n)$ is defined as the space of $m$-dimensional planes going through the origin in $n$-dimensional space, and thus may also be realized as an $m\times n$ matrix, this time modulo $GL(4,\mathbb{C})$ transformations.

To summarize, in the planar limit (color-ordered, super-)amplitudes in $\cN=4$ SYM only depend on the particle number $n$, the helicity degree $k$, $3n-15$ variables in the space of dual conformally invariant kinematics, and the order $L$ of loops or perturbative corrections.

\section{Multiple Polylogarithms and Symbols}\label{sec:MPLs}
Having specified the quantum numbers and kinematic variables that $A^{(L)}_{n,k}$ depends on, let us now move on to describe the class of functions it belongs to. A great deal of evidence from all explicit calculations to date, as well as from an analysis of the integrand \cite{ArkaniHamed:2012nw}  , suggests that at least for $k=0,1$, $A_{n,k}^{(L)}$ can be expressed in terms of Goncharov or \emph{multiple polylogarithms} (MPL) \cite{Chen,G91b,Goncharov:1998kja} (see also the review~\cite{Duhr:2014woa}) of weight $m=2L$. A function $F_m$ is defined to be an MPL
of weight $m$ if its total differential obeys
\be 
dF_m = \sum_{\phi_\beta\in{\Phi}} F_{m-1}^{\phi_\beta}\,\, d\log \phi_\beta\,,
\label{dFPhi}
\ee
such that that $F^{\phi_\alpha}_{m-1}$ is an MPL of weight $m-1$,
\be
dF_{m-1}^{\phi_\beta} = \sum_{\phi_\alpha\in{\Phi}} F_{m-2}^{\phi_\alpha,\phi_\beta} d\log \phi_\alpha\label{dFPhiCop}\,,
\ee
and so on, with the recursive definition terminating with the usual logarithms ($m=1$) on the left-hand side, and rational numbers ($m=0$) as coefficients of the total differentials on the right-hand side. The set
$\Phi$ of arguments of the dlogs is called the \emph{symbol alphabet}, and it encodes the positions of the possible branch points of $F$.
This iterative structure is part of the \emph{coaction} operation $\Delta$ \cite{Gonch3,Gonch2,Brown:2011ik,Duhr:2012fh} (sometimes loosely
referred to as a \emph{coproduct}), which `decomposes' an MPL
of weight $m$ to linear combinations of pairs of MPLs with weight $\{m-m_1,m_1\}$
for $m_1=0,1,\ldots m$.

In particular, the total differential \eqref{dFPhi} is essentially equivalent to the $\{m-1,1\}$ component of $\Delta$,
\be\label{eq:Deltan1}
\Delta_{m-1,1} F_m = \sum_{\phi_{\beta} \in \Phi} F_{m-1}^{\phi_\beta}\otimes
\big[\log\phi_{\beta}\mod\,(i\pi)\big]\,.
\ee
The coaction may be repeatedly applied to either the first or the second factor of the pair when $m_1>1$, yielding a further decomposition. As a result of the
coassociativity of the coaction there is a unique decomposition of an MPL of weight $m$ into subspaces of MPLs with weight $\{m_1,\ldots,m_r\}$, such that $\sum_{i=1}^r m_i=m$. Denoting the projection of the coaction on each of these subspaces by $\Delta_{m_1,\ldots,m_r}$, then equations \eqref{dFPhi} and \eqref{dFPhiCop} combine to yield the $\{m-2,1,1\}$ coproduct,
\begin{align}
\Delta_{m-2,1,1}F_m &= \sum_{\phi_\alpha,\phi_\beta\in{\Phi}} F_{m-2}^{\phi_\alpha,\phi_\beta} \otimes \log \phi_\alpha \otimes \log \phi_\beta\,,\label{Delta11}
\end{align}
where here, and in what follows, identification of $\log \phi$ factors up to $i\pi$ is implied. Furthermore, maximally iterating the procedure we just described defines the  \emph{symbol} \cite{2009arXiv0908.2238G,Goncharov:2010jf},
\be\label{symbol}
S[F_m]=\Delta_{\fwboxL{27pt}{{\underbrace{1,\ldots,1}_{m\,\,\text{times}}}}}F_m=\sum_{\phi_{\alpha_1},\ldots,\phi_{\alpha_m}} F_0^{\phi_{\alpha_1},\ldots,\phi_{\alpha_n}}\,  \left[\log {\phi_{\alpha_1}} \otimes\cdots \otimes\log \phi_{\alpha_m}\right]\,,
\ee
where one typically also simplifies the notation by replacing $\log \phi_{\alpha_i}\to \phi_{\alpha_i}$ for compactness. 

As comparison between \eqref{dFPhi} and \eqref{eq:Deltan1} reveals, derivatives only act on the rightmost factor of the coaction, and the same carries over to the symbol. Similarly, it can be shown that the singularities or discontinuities of MPLs are encoded in the leftmost factor of their coaction. Particularly at symbol level, the discontinuity of $F_m$ when going around a potential branch point $\phi_\beta=0$, is given by
\be\label{symbolDisc}
S[\text{Disc}_{\phi_\beta} (F_m)]=2\pi i \sum_{\phi_{\alpha_1},\ldots,\phi_{\alpha_m}} F_0^{\phi_{\alpha_1},\ldots,\phi_{\alpha_n}}\,  \delta_{\alpha_1\beta }\left[ {\phi_{\alpha_2}} \otimes\cdots \otimes \phi_{\alpha_m}\right]\,,
\ee
namely it is equivalent to clipping off the first entry.

\section{Symbol Alphabets and Cluster Algebras}

Once we have identified MPLs as the class of functions that contains $A^{(L)}_{n,k}$, the next step is to clarify how the symbol alphabet depends on the kinematic variables. For $n=6$, this was achieved by an explicit two-loop computation~\cite{DelDuca:2009au,DelDuca:2010zg}, as well as by the analysis of closely related integrals~\cite{DelDuca:2011ne,Dixon:2011ng}. For general $n$, strong motivation comes from the cluster algebra structure~\cite{Golden:2013xva} of the space of kinematics.

More precisely, we already mentioned in section \ref{sec:SymKin} that the space of momentum twistor kinematics can be realized as the quotient $Gr(4,n)/(\mathbb{C}^{*})^{n-1}$ of a Gra\ss mannian. The latter space is naturally endowed with a cluster algebra structure, so it is sensible to explore any implications it may have on the symbol alphabet.

There already exist many excellent introductions to cluster algebras \cite{2008arXiv0807.1960K,Lampe13,Fomin2016,Fomin2017}, as well as articles with detailed review sections on their relation to scattering amplitudes \cite{Vergu:2015svm,Drummond:2018dfd}, so we will not attempt to repeat this here. Instead, we will briefly highlight their main features, and outline a simple example of a Gra\ss mannian cluster algebra, which is relevant for our discussion.

Cluster algebras \cite{1021.16017,1054.17024,CAIII,CAIV} are commutative algebras equipped with a distinguished set of generators $a_i$, the \emph{cluster variables}, grouped into overlapping subsets $\{a_1,\ldots,a_d\}$ of \emph{rank} $d$,  the \emph{clusters}. Starting from an initial cluster, they may be constructed recursively by a \emph{mutation} operation on the cluster variables. They may also be generalized to contain \emph{frozen variables} or \emph{coefficients} $\{a_{d+1},\ldots,a_{d+m}\}$, whose main difference from the cluster variables is that they do not mutate.

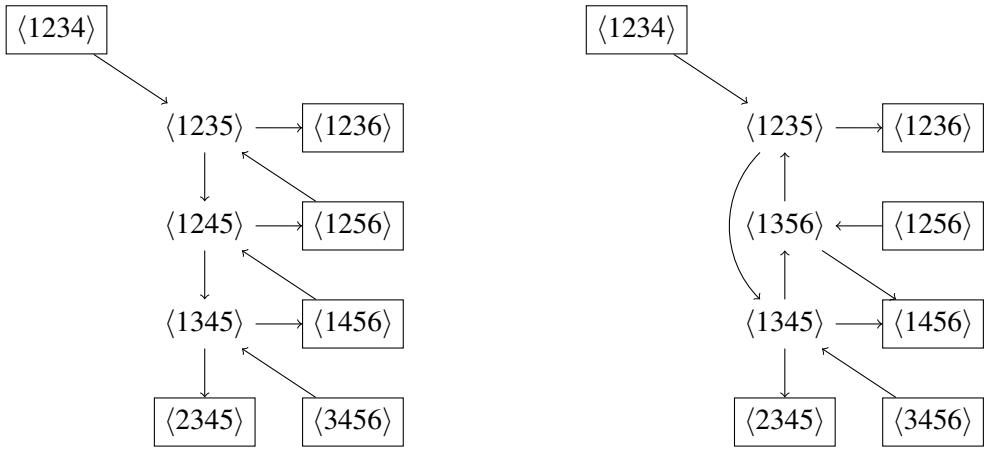
\begin{figure}
	\begin{minipage}{0.5\textwidth}
	\begin{center}
		\begin{tikzpicture}[scale=1.3]
		\node at (-1.5,4) [rectangle,draw] (a) {$\ab{1234}$};
		\node at (0,3) (b) {$\ab{1235}$};
		\node at (0,2) (c) {$\ab{1245}$};
		\node at (0,1) (b31) {$\ab{1345}$};
		\node at (0,0) [rectangle,draw] (b41) {$\ab{2345}$};
		\node at (1.5,3) [rectangle,draw] (b12) {$\ab{1236}$};
		\node at (1.5,2) [rectangle,draw] (b22) {$\ab{1256}$};
		\node at (1.5,1) [rectangle,draw] (b32) {$\ab{1456}$};
		\node at (1.5,0) [rectangle,draw] (b42) {$\ab{3456}$};
		\draw[->](a)--(b) ;
		\draw[->](b)--(c);
		\draw[->](c)--(b31);
		\draw[->](b31)--(b41);
		\draw[->](b32)--(c);
		\draw[->](c)--(b22);
		\draw[->](b22)--(b);
		\draw[->](b)--(b12);
		\draw[->](b31)--(b32);
		\draw[->](b42)--(b31);
	\end{tikzpicture}
		\end{center}
		\end{minipage}
		\begin{minipage}{0.5\textwidth}
		\begin{center}
			\begin{tikzpicture}[scale=1.3]
		\node at (-1.5,4) [rectangle,draw] (a) {$\ab{1234}$};
		\node at (0,3) (b) {$\ab{1235}$};
		\node at (0,2) (c) {$\ab{1356}$};
		\node at (0,1) (b31) {$\ab{1345}$};
		\node at (0,0) [rectangle,draw] (b41) {$\ab{2345}$};
		\node at (1.5,3) [rectangle,draw] (b12) {$\ab{1236}$};
		\node at (1.5,2) [rectangle,draw] (b22) {$\ab{1256}$};
		\node at (1.5,1) [rectangle,draw] (b32) {$\ab{1456}$};
		\node at (1.5,0) [rectangle,draw] (b42) {$\ab{3456}$};
		\draw[->](a)--(b) ;
		\draw[->](c)--(b);
		\draw[->](b31)--(c);
		\draw[->](b31)--(b41);
		\draw[->](c)--(b32);
		\draw[->](b22)--(c);
		\draw[->](b)--(b12);
		\draw[->](b31)--(b32);
		\draw[->](b42)--(b31);
		\draw[->](b) to[bend right=45]  (b31);
	\end{tikzpicture}
			\end{center}
			\end{minipage}
	\caption{Left: The quiver diagram for the $Gr(4,6)$ initial cluster. Right: The quiver that arises by mutating $\ab{1245}$ of the initial cluster, where the effect of the mutation is described in eqs.~\eqref{eq:mutation} and \eqref{eq:mutation1245} for the variable, and below eq.~\eqref{eq:PluckerRel} for the arrows of the quiver.}
\label{Gr46initial}
\end{figure}

For our purposes, cluster algebras can be described by directed graphs or quivers. In figure \ref{Gr46initial}, we depict the initial cluster of the $Gr(4,6)$ cluster algebra, relevant for six-particle scattering. Cluster and frozen variables correspond to the unboxed and boxed vertices of the graph, and we observe that they are all four-brackets, or equivalently Pl\"ucker coordinates, namely $4\times 4$ minors of the $4\times n$ matrix realization of $Gr(4,n)$, here for $n=6$. Nevertheless, frozen and cluster variables also have an important difference: The former are always of the form $\ab{i,i+1,i+2,i+3}$, modulo $n+i\sim i$ identifications, whereas this is not the case for the latter.

As we mentioned, mutations only act on cluster variables. Letting $a_k$ be a cluster variable, the arrows of a cluster containing it encode the information of how it will transform under mutation as follows,
\be\label{eq:mutation}
a_k\to a^{\prime}_k=\frac{1}{a_k}\left(\prod_{\textrm{arrows }i\to k}a_i+\prod_{\textrm{arrows }k\to j}a_j\right)\,.
\ee

For example, mutating $\ab{1245}$ in the left of figure \ref{Gr46initial}, we obtain
\be\label{eq:mutation1245}
\ab{1245}\to \frac{\ab{1235}\ab{1456}+\ab{1345}\ab{1256}}{\ab{1245}}=\ab{1356}\,,
\ee
where the last equality is obtained by means of the following Pl\"ucker relation\footnote{This identity is equivalent to the Schouten identity, as can be seen by the duality between $Gr(4,6)$ and $Gr(2,6)$, which replaces a four-bracket with a spinor bracket of its complement, e.g. $\ab{1234}\to \ab{56}.$}\,,
\be\label{eq:PluckerRel}
\ab{cdef}\ab{abef}-\ab{bdef}\ab{acef}+\ab{adef}\ab{bcef}=0.
\ee
for $e=1, f=5, a=2, b=3, c=4$ and $d=6$.

Apart from $a_k\to a^{\prime}_k$, in the new quiver resulting from this mutation, all the rest of the cluster variables and coefficients remain unchanged. However, the arrows connecting them are obtained by modifying those of the cluster before the mutation of $a_k$ as follows:
\begin{itemize}
\item For each path $i\to k\to j$ add an arrow $i\to j$, except if both $i$ and $j$ are frozen variables.
\item Reverse the direction of all arrows pointing to or originating from $k$.
\item Remove any pairs of arrows pointing in opposite directions, $\rightleftarrows$.
\end{itemize}
In this manner we see, for example, that the mutation of $\ab{1245}$ we performed on the $Gr(4,6)$ initial cluster, leads to the new cluster depicted on the right of figure \ref{Gr46initial}.

\begin{figure}
	\begin{center}
		\begin{tikzpicture}[scale=1.3]
		\node at (-1.5,4) [rectangle,draw] (a) {$\ab{1234}$};
		\node at (0,3) (b11) {$\ab{1235}$};
		\node at (0,2) (b21) {$\ab{1245}$};
		\node at (0,1) (b31) {$\ab{1345}$};
		\node at (0,0) [rectangle,draw] (b41) {$\ab{2345}$};
		\node at (1.5,3)  (b12) {$\ab{1236}$};
		\node at (1.5,2)  (b22) {$\ab{1256}$};
		\node at (1.5,1)  (b32) {$\ab{1456}$};
		\node at (1.5,0) [rectangle,draw] (b42) {$\ab{3456}$};
		\node at (3,3)  (b13) {\ldots};
		\node at (3,2)  (b23) {\ldots};
		\node at (3,1)  (b33) {\ldots};
		\node at (3,0)  (b43) {\ldots};
		\node at (5,3)  (b14) {$\ab{123\scalebox{0.9}{$n-1 $}}$};
		\node at (5,2)  (b24) {$\ab{12\, \scalebox{0.9}{$n-2 $}\,\scalebox{0.9}{$n-1 $}}$};
		\node at (5,1)  (b34) {$\ab{1\, \scalebox{0.9}{$n-3 $}\,\scalebox{0.9}{$n-2 $}\,\scalebox{0.9}{$n-1 $}}$};
		\node at (5,0) [rectangle,draw] (b44) {$\ab{\scalebox{0.9}{$n-4 $}\,\scalebox{0.9}{$n-3 $}\,\scalebox{0.9}{$n-2 $}\,\scalebox{0.9}{$n-1 $}}$};
		\node at (8,3) [rectangle,draw] (b15) {$\ab{123\scalebox{0.9}{$n $}}$};
		\node at (8,2) [rectangle,draw] (b25) {$\ab{12\, \scalebox{0.9}{$n-1 $}\,\scalebox{0.9}{$n $}}$};
		\node at (8,1) [rectangle,draw] (b35) {$\ab{1\, \scalebox{0.9}{$n-2 $}\,\scalebox{0.9}{$n-1 $}\,\scalebox{0.9}{$n $}}$};
		\node at (8,0) [rectangle,draw] (b45) {$\ab{\scalebox{0.9}{$n-3 $}\,\scalebox{0.9}{$n-2 $}\,\scalebox{0.9}{$n-1 $}\,\scalebox{0.9}{$n $}}$};
		\draw[->](a)--(b11) ;
		\draw[->](b11)--(b21);
		\draw[->](b21)--(b31);
		\draw[->](b31)--(b41);
		\draw[->](b11)--(b12);
	    \draw[->](b21)--(b22);
	    	\draw[->](b31)--(b32);
        	\draw[->](b22)--(b11);
		\draw[->](b32)--(b21);
		\draw[->](b42)--(b31);
		\draw[->](b12)--(b22);
		\draw[->](b22)--(b32);
		\draw[->](b32)--(b42);
		\draw[->](b14)--(b24);
		\draw[->](b24)--(b34);
		\draw[->](b34)--(b44);
		\draw[->](b14)--(b15);
	    \draw[->](b24)--(b25);
	    	\draw[->](b34)--(b35);
        	\draw[->](b25)--(b14);
		\draw[->](b35)--(b24);
		\draw[->](b45)--(b34);

	\end{tikzpicture}
		\end{center}
	\caption{Quiver diagram for the initial $Gr(4,n)$ cluster.}
\label{Gr4Ninitial}
\end{figure}
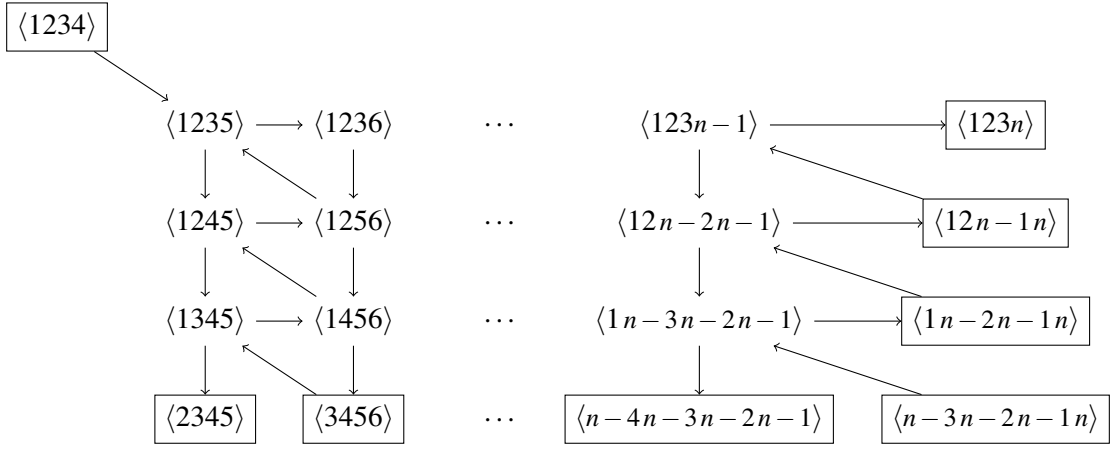

We have thus specified all the rules of the game, and obtaining the entire cluster algebra is just a matter of applying them over and over at each vertex of every quiver we encounter. For $Gr(4,n)$ with $n=6,7$, relevant for scattering amplitudes with the same particle multiplicity $n$, we see that after a certain number of mutations, we return to a cluster we had encountered before. Namely the cluster algebra is finite.\footnote{In the Dynkin classification of finite cluster algebras, $Gr(4,6)$ and $Gr(4,7)$ can thus be shown to correspond to the $A_3$ and $E_6$ cluster algebras (with coefficients), respectively.} The initial quiver for generic $Gr(4,n)$, from which the $n=7$ case may be studied with the same set of rules we spelled out, is depicted in figure~\ref{Gr4Ninitial}.

Specifically, for $Gr(4,6)$ ($Gr(4,7)$), one finds 9 (42) different cluster variables, distributed in 14 (833) distinct clusters (the order of the variables of each cluster does not matter). Quite remarkably, in \cite{Golden:2013xva} these cluster variables were found to precisely agree with the symbol alphabet of the then known six- and seven-particle amplitudes, thus lending support to the expectation that the same should hold true to all loops. 

More precisely, the cluster variables are homogeneous polynomials of four-brackets, and to obtain the symbol alphabet one needs to consider a complete set of multiplicatively independent, scale-invariant combinations thereof, also including the frozen variables.

For the six-particle amplitude, or equivalently $Gr(4,6)$, a convenient such choice for the symbol alphabet is~\cite{Caron-Huot:2019bsq}\footnote{Note that any set of equal size, consisting of multiplicatively independent combinations of these letters, would make an equally valid choice. Indeed, in the original literature~\cite{Dixon:2011pw,Dixon:2011nj,Dixon:2013eka,Dixon:2014voa,Dixon:2014iba} the letters $u_i$ and $1-u_i$ were used. The relation with the presently used alphabet is $a_i=u_i/(u_{i-1} u_{i+1})$ and $m_i=1-1/u_i$.}
\be\label{eq:hexletters}
a_1=\frac{\textcolor{blue}{\ab{1245}}^2\ab{3456}^2\ab{6123}^2}{\ab{1234}\ab{2345}\ldots \ab{6123}}\,,\quad m_1=\frac{\textcolor{blue}{\ab{1356}\ab{2346}}}{\ab{1236}\ab{3456}}\,,\quad y_1=\frac{\textcolor{blue}{\ab{1345}\ab{2456}}\ab{1236}}{\textcolor{blue}{\ab{1235}\ab{1246}}\ab{3456}}\,,
\ee
as well as two more cyclic transformations $l_1\to l_{1+i}$ with $l \in \{a,m,y\}$ induced by shifting $Z_m\to Z_{m-2i}$ on the right-hand side. The cluster variables, also known as $\cA$-coordinates, are color-coded in blue.

For the seven-particle amplitude, or $Gr(4,7)$, a choice for the corresponding symbol alphabet is~\cite{Drummond:2014ffa}
\begin{align}
\label{eq:heptagonletters}
a_{11} &=
\frac{\ab{1234} \ab{1567} \textcolor{blue}{\ab{2367}}}{\ab{1237} \ab{1267} \ab{3456}}\,,
&
a_{41} &=
\frac{\textcolor{blue}{\ab{2457}} \ab{3456}}{\ab{2345} \ab{4567}}\,,
\nonumber
\\
a_{21} &=
\frac{\ab{1234} \textcolor{blue}{\ab{2567}}}{\ab{1267} \ab{2345}}\,,
&
a_{51} &=
\frac{\textcolor{blue}{\ab{1(23)(45)(67)}}}{\ab{1234} \ab{1567}}\,,
\\
a_{31} &=
\frac{\ab{1567} \textcolor{blue}{\ab{2347}}}{\ab{1237} \ab{4567}}\,,
&
a_{61} &=
\frac{\textcolor{blue}{\ab{1(34)(56)(72)}}}{\ab{1234} \ab{1567}}\,,
\nonumber
\end{align}
where we have again denoted the cluster $\cA$-coordinates in blue, and
\be\label{brbilinear}
\ab{a(bc)(de)(fg)} \equiv
\ab{abde} \ab{acfg} - \ab{abfg} \ab{acde}\,,
\ee
together with $a_{ij}$ obtained from $a_{i1}$ by cyclically relabeling the momentum twistors $Z_m \to Z_{m + j - 1}$. It is interesting to note that for $n=7$, and more generally when $n$ is odd, any cluster $\cA$-coordinate can be rendered invariant under rescalings $Z_i\to t Z_i$ by suitable products of powers of frozen variables.

\section{Hexagon and Heptagon Functions}\label{sec:HexHepFuns}
In the previous sections, we described the essential characteristics of the function spaces containing six- and seven-particle amplitudes in $\cN=4$ SYM theory. The virtue of the bootstrap method is that it evades Feynman diagrams by first constructing these spaces, and then pinning down the amplitude by comparing with certain kinematic limits, where independent information for the latter can be obtained. Here, we will briefly describe how these function spaces are constructed, including certain additional analytic properties they obey.

As a consequence of locality,  amplitudes can only have singularities when  intermediate particles go on-shell. Particularly for massless planar theories in the Euclidean region, this happens if and only if the Mandelstam invariants \eqref{eq:Mandelstamtox}, or equivalently four-brackets $\ab{i i-1 j j-1}$ become zero or infinite. Given that the singularities of multiple polylogarithms are encoded in the first entry of their symbols, and that most of the letters in \eqref{eq:hexletters}-\eqref{eq:heptagonletters} also contain different types of four-brackets, this implies the \emph{first entry condition}~\cite{Gaiotto:2011dt}: 
\be\label{eq:firstentries}
\text{First symbol entry of}\,\,\, A_{n,k}\in
\begin{cases}
a_i\,,\,\, i=1,\ldots, 3\,,& \text{for }n=6\,,\vspace{12pt}\\
a_{1i}\,,\,\, i=1,\ldots, 7\,,& \text{for }n=7\,.
\end{cases}
\ee

Not every sequence of letters of a given alphabet yields a symbol that corresponds to a function, as the latter also has the property that its double derivatives with respect to two different independent variables 
commute. Namely the function $F$ (as well as its coproducts) obeys
\be
\frac{\partial^2F}{\partial x_i\partial x_j}\ =\
\frac{\partial^2F}{\partial x_j\partial x_i} \,, \qquad i \neq j,
\label{eq:commdoublederiv}
\ee
where particularly for $A_{n,k}$ the independent variables $x_i, x_j$ can be any subset of the $3n-15$ independent cross ratios, or of the $x$-coordinates provided in the appendix for $n=6,7$. This condition, when computed using eqs.~\eqref{dFPhi} and~\eqref{dFPhiCop}, induces linear relations
\be
\sum_{\alpha,\beta=1}^{|\Phi|} D_{i\alpha\beta} \, F^{\phi_{\alpha},\phi_{\beta}}=0\,,\qquad i=1,2,\ldots,l\,,\label{DoubleCopMatrix}
\ee
between the double coproducts $F^{\phi_\alpha,\phi_\beta}$, where $|\Phi|$ denotes the size of the alphabet, known as the $\{m-2,1,1\}$ \emph{integrability conditions}. As was discussed in \cite{Dixon:2016nkn}, there exist $l=26$ equations for the $9^2=81$ double coproducts when $n=6$, and $l=729$ equations for the $42^2=1764$ double coproducts when $n=7$. The matrix $D$ is purely numeric, i.e. independent of the kinematics, and is provided for $n=6,7$ as an ancillary file accompanying the \texttt{arXiv} submission of this contribution.

These equations, allow us to recursively construct the space of weight-$m$ functions built out of the $n$-particle alphabet with $n=6,7$, which we shall denote $\cH_{n,m}$, from $\cH_{n,m-1}$, as follows: Given a basis of functions on $\cH_{n,m-1}$, consider their $\{m-2,1\}$ coproduct representation \eqref{eq:Deltan1}, and attach another letter to them to the right in all possible ways. In this $\{m-2,1,1\}$ tensor product space that resembles eq.~\eqref{Delta11}, $\cH_{n,m}$ is obtained by imposing \eqref{DoubleCopMatrix} on its elements. Thus the bootstrap procedure transforms amplitude computations to linear algebra, and in the past software relying on finite field methods, such as \texttt{IML} \cite{Chen:2005:BBC:1073884.1073899}, \texttt{SageMath} \cite{SageMath} or \texttt{SpaSM} \cite{spasm}, has been used in order to solve the resulting large systems of linear equations efficiently. More recently, packages or frameworks with dedicated capabilities for constructing integrable symbols,   such as \texttt{SymBuild} \cite{Mitev:2018kie} or \texttt{FiniteFlow} \cite{Peraro:2019svx}, have also appeared.

The procedure we described works equally well for functions or for their symbols, and guarantees that if $\cH_{n,m-1}$ has physical branch cuts, so will $\cH_{n,m}$, with one type of exception for the case of functions: The space of solutions of the integrability conditions \eqref{DoubleCopMatrix} also contains functions such as
\be\label{eq:BadBranchFunction}
\zeta_{m-1}\log \phi_\alpha
\ee
where $\zeta_{m-1}$ is the Riemann zeta function and $\phi_\alpha$ is a letter that cannot appear as a first entry \eqref{eq:firstentries}, which do not have physical branch cuts.

We thus have to eliminate such functions from our space, and one way to do so is by noting that the branch point at $\phi_\alpha=0$ also manifests itself as a pole in the derivative of the function. So we may ensure that the function is analytic at $\phi_\alpha=0$ by imposing that the corresponding residue of its derivative, or equivalently its left coproduct factor, due to \eqref{dFPhi}, vanishes as $\phi_\alpha\to 0$.

In practice, it is more convenient to impose such \emph{branch cut conditions} on kinematic limits where more letters vanish simultaneously. Here we will only mention one of the limits that has been used in the literature for $n=6$, the soft or multi-Regge limit. For the six-particle alphabet \eqref{eq:hexletters}, in a particular orientation it amounts to\footnote{In the more conventional cross ratios \eqref{eq:udef}, this limit is equivalent to $u_1\to 1, u_2,u_3\to0$ with $u_i/(1-u_1)$ held fixed for $i=2,3$.}
\be\label{eq:soft1}
\text{soft}_1: \,\,a_1\to \infty\,,\,\, a_3\to \frac{1}{a_2}\,,\,\, m_2\to \frac{\sqrt{a_1}}{\sqrt{a_2}}\,,\,\, m_3\to \sqrt{a_1}{\sqrt{a_2}}\,,\,\, y_1\to 1\,,\,\,\,\text{with}\,\,\,a_2, y_2, y_3, {m_1}{\sqrt{a_1}}   \,\,\,\text{fixed}\,,
\ee
and in this limit functions in $\cH_{6,m}$ should additionally obey \cite{Dixon:2013eka,Dixon:2015iva}
\be\label{eq:hexBranchCuts}
F^{m_1}\big|_{\text{soft}_1}=F^{y_3}\big|_{\text{soft}_1}=F^{y_2}\big|_{\text{soft}_1}=0\,.
\ee
We also obtain two more limits and corresponding branch cut conditions by cyclic permutation.

To reiterate, at this stage we may define the space of \emph{hexagon or heptagon symbols} as the set of symbols made of the alphabets \eqref{eq:hexletters} or \eqref{eq:heptagonletters} and first entry as dictated by eq.~\eqref{eq:firstentries}, that are constructed recursively in weight by imposing the integrability conditions \eqref{eq:commdoublederiv}. The corresponding space of \emph{$n$-gon functions} at weight $m$, $\cH_{n,m}$, may be defined by additionally imposing branch cut conditions such as eq.~\eqref{eq:hexBranchCuts} at each weight for $n=6$. Analogous relations may also be derived for $n=7$~\cite{HenkePapathanasiou,DixonLiu}.

\section{Extended Steinmann Relations \& Cluster Adjacency}\label{sec:SteinClus}

The space of functions $\cH_{n,2L}$ we have constructed so far is certainly large enough to encompass the $L$-loop $n$-particle amplitude for $n=6,7$, and it was indeed used at the initial stages of the bootstrap~\cite{Dixon:2011pw,Dixon:2011nj,Dixon:2013eka,Dixon:2014voa,Dixon:2014iba,Drummond:2014ffa}. Nevertheless, it is also largely redundant, in the sense that it additionally contains functions that need never appear in the amplitude or its derivatives. Here, we will focus on one of the main discoveries reported in this contribution, the extended Steinmann relations or cluster adjacency, which drastically simplify the construction of $\cH_{n,2L}$ by eliminating all such redundancies visible at symbol level.

\begin{figure}
\begin{center}
\includegraphics[width=0.7\textwidth]{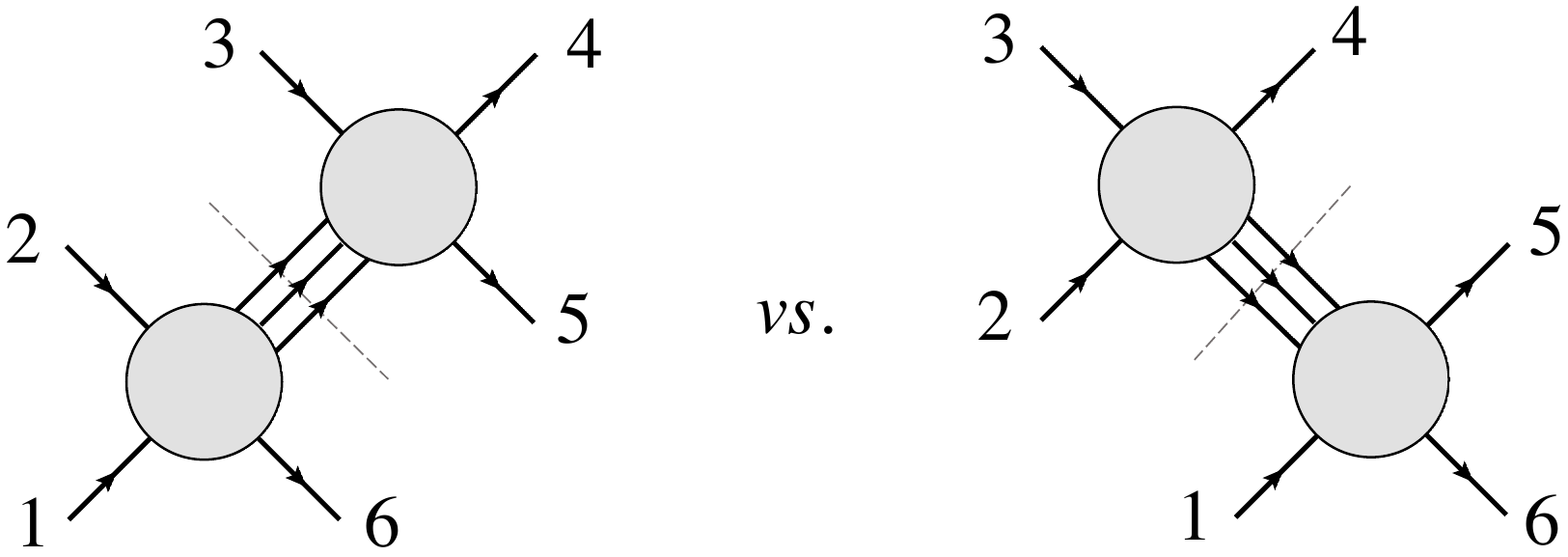}
\end{center}
\caption{Illustration of the channels $s_{345}$ and $s_{234}$ for $3\to3$
kinematics. The discontinuity in one channel should not know about
the discontinuity in the other channel.}
\label{fig:stein}
\end{figure}

To this end, it will be instructive to first review  the usual Steinmann relations~\cite{Steinmann,Steinmann2,Cahill:1973qp}, which demand that the double discontinuities of any Feynman diagram (and thus of the amplitude they contribute to) vanish when taken in overlapping channels. As we've seen in section \ref{sec:MPLs}, a discontinuity is labeled by the Mandelstam invariant $s_{i,\ldots,j-1}$ which is analytically continued around its branch point. By virtue of the Cutkosky rules \cite{Cutkosky:1960sp}, at the same time this discontinuity is obtained by placing on-shell the internal particles whose total energy equals $s_{i,\ldots,j-1}$, namely by replacing their propagators with delta functions. This is the notion of a cut, which splits the Feynman diagram into two parts, as seen in figure \ref{fig:stein}.

In the same vein, overlapping channels correspond to cut lines that intersect, namely they divide the external particles of the Feynman diagram into four non-empty sets. In the example of the figure, these sets are $\{2\}$, $\{3,4\}$, $\{5\}$, and $\{6,1\}$. Focusing on three-particle Mandelstam invariants, but allowing the number of external particles $n$ to be arbitrary, the Steinmann relations at the level of the amplitude then imply
\be
\label{eq:disc_SteinmannA}
  \Disc_{s_{j,j+1,j+2}}\left(\Disc_{s_{i,i+1,i+2}}\left({A}_{n,k}\right) \right)=0\,,\quad \text{for}\,\,j=i\pm 1,i\pm 2\,,
\ee
with obvious generalizations to higher-particle Mandelstam invariants as well. In particular, we will not be considering two-particle invariants, since it is necessary for an invariant to be independent for the sake of analytic continuation. That is, no other Mandelstam invariant is allowed to change sign but the one we analytically continue, and this is generically not the case with two-particle invariants.

As eq.~\eqref{symbolDisc} reveals, at the level of the symbol a discontinuity around $\phi_\beta=0$ amounts to clipping off this particular letter from its first entry, with iterated discontinuities obtained by applying this procedure repeatedly. Therefore the Steinmann relations are statements about which letters can appear next to each other in the first two entries of the symbol. Inspecting the $n=6,7$ letters \eqref{eq:hexletters}-\eqref{eq:heptagonletters}, we see that only $a_i$ and $a_{1i}$ are proportional to three-particle invariants. Therefore, letting $F$ denote the appropriately normalized amplitudes, or any of their coproducts, or any finite integral within the space of MPLs with these alphabets, eq.~\eqref{eq:disc_SteinmannA} may be recast as~\cite{Caron-Huot:2016owq,Dixon:2016nkn}
\be\label{eq:FabSteinw2}
\text{Steinmann Relations:}
\left\{\begin{array}{ll}
F^{a_i,a_{i+1}}=0\,,1\le i\le 3\,, &\text{for}\,\,n=6\\
F^{a_{1i},a_{1i+\delta}}=0\,, \,\delta=1,2\,,\, 1\le i\le7\,,&\text{for}\,\,n=7 
\end{array}\right\}
\text{if $F$ function of weight 2}\,.
\ee

Remarkably, analysis of a wealth of data obtained by the amplitude bootstrap, reveals that these restrictions on adjacent pairs of symbol letters apply not only in the first two slots, but to \emph{all} depths in the symbol. In other words, the bootstrap points, for the first time, to a new property we will refer to as the \emph{extended Steinmann relations}~\cite{Caron-Huot:2019bsq}. For the case at hand, it reads that for any amplitude (coproduct) or Feynman integral $F$ within the hexagon or heptagon function space $\cH_{n,m}$, \footnote{These results for $n=6$ were initially reported at {\it Amplitudes 2017}, in a talk by one of the authors~\cite{YorgosAmps17}.}
\be\label{eq:FabExtStein}
\boxed{\text{Extended Steinmann Relations:}
\left\{\begin{array}{ll}
F^{a_i,a_{i+1}}=0\,,1\le i\le 3\,, &\text{for}\,\,n=6\,.\\
F^{a_{1i},a_{1i+\delta}}=0\,, \,\delta=1,2\,,\, 1\le i\le7\,,&\text{for}\,\,n=7 \,.
\end{array}\right.}
\ee
In other words, all physical, integrable functions within $\cH_{n,m}$ obey\footnote{Note that the integrability conditions in combination with \eqref{eq:FabSteinw2} automatically imply that the latter equations hold also when the order of letters is reversed, so we do not need to consider them separately.} an additional 3 conditions for $n=3$, and 14 for $n=7$, on their double coproducts. In tables \ref{tab:HexSymbolDim} and \ref{tab:HepSymbolDim}, we provide the dimension of $\cH_{n,m}$ (first line), as well as the dimensions of its subspaces additionally satisfying the Steinmann (second lines) or extended Steinmann (third line) relations, eqs.~\eqref{eq:FabSteinw2} and~\eqref{eq:FabExtStein}, respectively. Clearly, the drastic reduction in the size of the latter subspace implies that it is far simpler to construct, yet still contains all the physical quantities we are interested in. From this point onwards, we will therefore incorporate the condition \eqref{eq:FabExtStein} in the construction of our space $\cH_{n,m}$, and denote the latter as the \emph{extended Steinmann $n$-gon space}.

\renewcommand{\arraystretch}{1.25}
\begin{table}[!t]
\centering
\resizebox{\textwidth}{!}{
\begin{tabular}[t]{l c c c c c c c c c c c c c c}
\hline\hline
weight $n$
& 0 & 1 & 2 & 3 & 4 &  5 &  6 &  7 &  8 &  9 & 10 & 11 & 12 & 13 \\
\hline\hline
First entry
& 1& 3 & 9 & 26 & 75 & 218 & 643 & 1929 & 5897 & ? & ? & ? & ? & ? 
\\\hline
Steinmann
& 1 & 3 & 6 & 13 & 29 & 63 & 134 & 277 & 562 & 1117 & 2192 & 4263 & 8240 & ?
\\\hline
Ext.~Stein.
& 1 & 3 & 6 & 13 & 26 & 51 & 98 & 184 & 340 & 613 & 1085 & 1887 & 3224 & 5431 
\\\hline\hline
\end{tabular}}
\caption{The dimensions of the hexagon, Steinmann hexagon, and extended Steinmann hexagon spaces at symbol level.}
\label{tab:HexSymbolDim}
\end{table}

\renewcommand{\arraystretch}{1.25}
\begin{table}[!t]
\centering
\begin{tabular}[t]{l c c c c c c c c}
\hline\hline
weight $n$ & 0 & 1 & 2 & 3 & 4 &  5 &  6 &  7  \\\hline\hline
First entry & 1 & 7 & 42 & 237 & 1288 & 6763 & ? & ?  \\\hline
Steinmann & 1 & 7 & 28 & 97 & 322 & 1030 & 3192 & 9570  \\\hline
Ext.~Stein. & 1 & 7 & 28 & 97 & 308 & 911 & 2555 & 6826 \\\hline\hline
\end{tabular}
\caption{The dimensions of the heptagon, Steinmann heptagon, and extended Steinmann heptagon  spaces at symbol level. }\label{tab:HepSymbolDim}
\end{table}

Remarkably, it has been observed that the extended Steinmann relations are equivalent to the following rule (under the assumption of the initial entry condition)~\cite{Drummond:2017ssj},
\vspace{6pt}

\fbox{\begin{minipage}{0.9\textwidth}{
\emph{Cluster adjacency: In a symbol whose alphabet contains $\textrm{Gr}(4,n)$ cluster $\cA$-coordinates, two of them can appear consecutively	only if there exists a cluster where they both appear.}}
\end{minipage}}

\vspace{6pt}

For every cluster $\cA$-coordinate, cluster adjacency thus gives rise to  the notion of its \emph{neighbor set} \cite{Drummond:2018dfd}, the union of all clusters containing the $\cA$-coordinate in question, which in other words contains all the other variables (including the frozen ones), that can appear next to it in the symbol.

For $n=6$, these neighbor sets are
\begin{align}
    \ns{\ab{1245}} &= \{\ab{1245}, \ab{2456}, \ab{1345}, \ab{1246},  \ab{1235},\, \text{\& frozen variables.}\}\,,\label{eq:ns1245}\\
        \ns{\ab{1235}} &= \{\ab{1235}, \ab{2456}, \ab{2356}, \ab{1356}, \ab{1345}, \ab{1245}, \,\text{\& frozen variables.}\}\,,\label{eq:ns1235}
\end{align}
as well as their cyclic permutations, two for the first line and five for the second.

As we've discussed, actual symbol letters are conformally invariant, namely products of the cluster and frozen variables that are invariant under rescalings of the twistors, i.e. homogeneous. Considering such products on the right-hand side of eqs.~\eqref{eq:ns1245}-\eqref{eq:ns1235} then defines the corresponding \emph{homogeneous neighbor sets},
\begin{align}
    \hns{\ab{1245}} &= \hns{a_1} = \{a_1,m_2,m_3,y_1, y_2 y_3\}\,,\label{eq:hns1245}\\
        \hns{\ab{1235}} &= \{a_1,a_2,\frac{m_1}{y_2},\frac{m_2}{y_2 y_3},m_3, y_1 y_2 y_3\}\,.\label{eq:hns1235}
\end{align}
again plus cyclic permutations. In the first line we could also promote the left-hand side to the conformally invariant letter $a_1$, since $\ab{1245}$ is the only cluster variable it depends on. This is not possible for the second line, relevant for the remaining letters $m_i,y_i$.

For $n=7$, all letters depend on a single $\cA$-coordinate. We can thus directly focus on  the homogeneous neighbor sets, which are generated by

  \begin{align}
    \hns{a_{11}} =\{ &a_{11}, a_{14}, a_{15}, a_{21}, a_{22}, a_{24}, a_{25}, a_{26}, a_{31}, a_{33}, a_{34}, a_{35}, a_{37}, a_{41},a_{43}, a_{46}, a_{51},\label{heptns}\\
     \qquad        &a_{53}, a_{56}, a_{62}, a_{67}\}\nonumber\\
    \hns{a_{21}} =\{&a_{11}, a_{13}, a_{14}, a_{15}, a_{17}, a_{21}, a_{23}, a_{24}, a_{25}, a_{26},a_{31}, a_{33}, a_{34}, a_{36},a_{37}, a_{41}, a_{43},\label{ns7first}\\
    \qquad        &a_{45}, a_{46}, a_{52}, a_{53}, a_{55}, a_{57}, a_{62}, a_{64}, a_{66}\}\nonumber\\
    \hns{a_{41}} =\{&a_{11}, a_{13}, a_{16}, a_{21}, a_{23}, a_{24}, a_{26}, a_{31}, a_{33}, a_{35}, a_{36}, a_{41}, a_{43}, a_{46},a_{51}, a_{62}, a_{67}\}\\
    \hns{a_{61}} =\{&a_{12}, a_{17}, a_{23}, a_{25}, a_{27}, a_{32}, a_{34}, a_{36}, a_{42}, a_{47}, a_{52}, a_{57}, a_{61}\}\,,\label{ns7last}
\end{align}
together with images under parity transformations, $a_{2i}\leftrightarrow a_{3,i-1}$ and  
$a_{4i}\leftrightarrow a_{5i}$, as well as cyclic permutations.

It is easy to check that the Steinmann relations \eqref{eq:FabSteinw2} and \eqref{eq:FabExtStein} per se, essentially correspond to eqs.~\eqref{eq:hns1245} and \eqref{heptns}, respectively. This also reveals that the equivalence between Steinmann relations and cluster adjacency at the level of the function space $\cH_{n,2L}$ is quite nontrivial: Namely, first entry, integrability and extended Steinmann relations automatically imply a number of additional double coproduct conditions (22 for $n=6$ and for 448 for $n=7$), which coincide with those predicted by the remaining cluster adjacent homogeneous neighbor sets, eqs.~\eqref{eq:hns1235} and~\eqref{ns7first}-\eqref{ns7last}. In total, double coproducts $F^{\phi_\alpha,\phi_\beta}$ of all functions $F\in \cH_{n,m}$ obey 41 (1191) linear relations for $n=6$ (7). Conversely, cluster adjacency provides some of these relations, which may also be thought of as implying some of the ``simplest'' integrability conditions.

Before closing this section, let us stress again that extended Steinmann/cluster adjacency holds not only for the entire amplitudes, but also for individual integrals \cite{Drummond:2017ssj,Caron-Huot:2018dsv}. Thus it may well be that the amplitude bootstrap reveals, for the first time, a universal property of quantum field theory, with far wider applicability. In $\cN=4$ SYM, this property has been observed also for the $n$-point MHV amplitude at one and two loops~\cite{Golden:2019kks}.

\section{Coaction Principle}\label{sec:Coaction}
\subsection{Amplitude Normalizations}
So far, we have mentioned that it is possible to factor out a universal infrared-divergent part $A_n^{\text{IR}}$ from the amplitude,
\be\label{eq:AtoAfin}
A_{n,k} =A_n^{\text{IR}} A_{n,k}^\text{fin}\,,
\ee
such that we can focus entirely on the finite, normalized amplitude $A_{n,k}^\text{fin}$. It is important to bear in mind, however, that this factorization is not unique: Similarly to the difference between renormalization schemes in any gauge theory, there is still freedom in adding certain finite terms to $A_n^{\text{IR}}$. A main thesis of our work is that it is meaningful to \emph{tune} the definition of this normalization factor, such that the remaining finite, normalized amplitude becomes simpler to compute, and manifests certain important physical and mathematical properties. The coaction principle we will discuss in this section, is precisely such a mathematical property.

Indeed, the strategy of tuning $A_n^{\text{IR}}$ has already proven fruitful once in the past. Initially, this factor was chosen to be the BDS ansatz~~\cite{Bern:2005iz}, which is essentially the exponentiated one-loop amplitude, times the cusp anomalous dimension $\Gcusp$,
\be
\frac14 \Gcusp(g^2)\ =\ g^2 - 2\,\zeta_2\,g^4 + 22\,\zeta_4\,g^6
- \Bigl[ 219\, \zeta_6 + 8 \, (\zeta_3)^2 \Bigr] \, g^8 + \cdots,
\label{Gcusp}
\ee
in turn known to all loops~\cite{Beisert:2006ez} thanks to the integrability of planar $\cN=4$ SYM, see for example the review \cite{Freyhult:2010kc} and references therein. However, neither the BDS ansatz, nor the respective normalized amplitude individually satisfy the Steinmann relations, even though the product of the two, namely the full amplitude, does.\footnote{This is a simple consequence of the fact that the 1-loop amplitude also has dependence on three-particle Mandelstam invariants, and of the non-distributivity of analytic continuation over multiplication. Note that the 1-loop amplitude per se does satisfy the Steinmann relations, but generic powers thereof, and thus also its exponential, do not.} This implies that the BDS-normalized amplitude is harder to bootstrap, since it is only contained in the larger space whose dimension is indicated in the first line of tables \ref{tab:HexSymbolDim} and \ref{tab:HepSymbolDim} for $n=6,7$, and whose construction is more challenging than the spaces of the lines below, due to its size.

Nevertheless, when $n$ is not a multiple of 4, it is possible to modify $A_n^{\text{IR}}$, such that the resulting $A_{n,k}^\text{fin}$ does obey the (extended) Steinmann relations, and hence is contained in the spaces of smaller dimension, depicted in the second (and third) line of tables \ref{tab:HexSymbolDim} and \ref{tab:HepSymbolDim}. This is known as the BDS-like ansatz~\cite{Alday:2009dv}, which first appeared as a natural choice for $A_n^{\text{IR}}$ at strong coupling. It is obtained from the BDS ansatz by removing all dependence on three-particle invariants in a unique conformally invariant manner, such that when replacing \eqref{eq:AtoAfin} in \eqref{eq:disc_SteinmannA}, the normalization factor commutes with the discontinuities.

For concreteness, we denote the BDS-like normalized amplitude as
\be
\cE_{n,k}\equiv\frac{A_{n,k}}{\,A^{(0)}_{n,0}\,A^{\text{BDS-like}}_{n}}\,,\quad n\,\text{mod}\, 4\ne 0\,,
\ee
where $A_n^{\text{IR}}\to A_n^{\text{BDS-like}}$, and it proves more convenient to additionally divide by the tree-level MHV superamplitude, $A^{(0)}_{n,0}$. The precise form of the BDS and BDS-like ans\"atze will not be important for our purposes (it may be found in the original references, or e.g. \cite{Alday:2008yw,Yang:2010as}), however their ratio is closely related to the 1-loop correction to $\cE_{n,0}$\,,
where
\be
\frac{A_n^{\text{BDS}}}{A_n^{\text{BDS-like}}}=\exp\left[\frac{\Gamma_{\text{cusp}}}{4} \cE^{(1)}_{n,0}\right],
\label{BDSlikeBDSn}
\ee
and, explicitly for $n=6,7$,
\bea
\cE^{(1)}_{6,0} &=& \, \sum_{i=3}^7  \Li_2\left(1-\frac{1}{u_i}\right)\,,\label{E61def}\\
\cE^{(1)}_{7,0} &=& \sum_{i=1}^7 \biggl[ \Li_2\left(1-\frac{1}{u_i}\right)
  + \frac{1}{2} \log \left(\frac{u_{i+2}u_{i{-}2}}{u_{i+3}u_{i}u_{i{-}3}}\right)
               \log u_i \biggr]\,.
\label{E71def}
\eea
With the help of \eqref{BDSlikeBDSn}, it is easy to convert between BDS and BDS-like normalizations. For example, the BDS-normalized MHV amplitude, originally expressed in terms of an exponentiated \emph{remainder function} $\cR_n$, is related to $\cE_{n,0}$ by
\be\label{eq:R6}
e^{\cR_n}=e^{-\frac{\Gamma_{\text{cusp}}}{4} \cE^{(1)}_{n,0}} \cE_{n,0}\,.
\ee

Given the drastic reduction of the relevant function space that was achieved when modifying the amplitude normalization from BDS to BDS-like, it is natural to ask whether any further modifications with similar effect can be made. Any such modification can no longer depend on the kinematics, as it would otherwise spoil the first entry condition, or the Steinmann relations.

Therefore the only potential freedom in modifying the amplitude normalization is by a coupling-dependent constant, $\rho(g^2)$. Indeed, as we will explain shortly, at least for $n=6$ it is advantageous to define a new, ``cosmic'' normalization \cite{Caron-Huot:2019vjl,Caron-Huot:2019bsq},
\be\label{eq:CosmicNorm}
\cE'_{6,k}\equiv\frac{\cE_{6,k}}{\rho(g^2)}\,,
\ee
such that the space containing $\cE'_{6,k}$ and its iterated derivatives contains the minimal amount of independent constants, and respects a coaction principle. In particular, the latter will be instrumental in determining the value of $\rho(g^2)$, which we quote here to 7 loops,
\bea
\rho(g^2) &=& 1 + 8 (\zeta_3)^2 \, g^6  - 160 \zeta_3 \zeta_5 \, g^8
+ \Bigl[ 1680 \zeta_3 \zeta_7 + 912 (\zeta_5)^2 - 32 \zeta_4 (\zeta_3)^2 \Bigr]
\, g^{10}
\nonumber\\
&&\null\hskip0.0cm
- \Bigl[ 18816 \zeta_3 \zeta_9 + 20832 \zeta_5 \zeta_7
  - 448 \zeta_4 \zeta_3 \zeta_5 - 400 \zeta_6 (\zeta_3)^2 \Bigr] \, g^{12}
\nonumber\\
&&\null\hskip0.0cm
+ \Bigl[ 221760 \zeta_3 \zeta_{11} + 247296 \zeta_5 \zeta_9 + 126240 (\zeta_7)^2
   - 3360 \zeta_4 \zeta_3 \zeta_7 - 1824 \zeta_4 (\zeta_5)^2
	\nonumber\\
&&\null\hskip0.7cm
 - 5440 \zeta_6 \zeta_3 \zeta_5 - 4480 \zeta_8 (\zeta_3)^2 \Bigr] \, g^{14}
\ +\ {\cal O}(g^{16}).
\label{rho}
\eea
When first introduced, the computation of the normalization factor \eqref{rho} was carried out order by order in perturbation theory, in parallel with the amplitude. Very recently, however, a finite-coupling conjecture for the latter has been made~\cite{Basso:2020xts}.

Note that $S[\rho]=1$, therefore at symbol level there is no difference between the BDS-like and cosmic normalizations. As the seven-particle bootstrap has only been carried out at symbol level to date, the status of its cosmic normalization is unclear (see however ref.~\cite{DixonLiu}). 


\subsection{The role of transcendental constants}
Before describing the coaction principle, it will be necessary to give some background on the inclusion of transcendental constants in the function spaces we are constructing. These constants generically arise when setting arguments of multiple polylogarithms, that our function spaces consist of, to specific values. For example, the Riemann zeta function for $m> 1$ may be obtained as an evaluation of the weight-$m$ classical polylogarithm $\text{Li}_m(z)$,
\be\label{eq:ZetaLi1}
\zeta_m=\text{Li}_m(1)\,.
\ee
For the purposes of this contribution, it will be sufficient to consider a class of transcendental constants that are simple generalizations of the Riemann zeta function, known as multiple zeta values (MZVs),
\be
\zeta_{m_1, \dots, m_l} \equiv \sum_{k_1 > \cdots > k_d > 0} \frac{1}{k_1^{m_1} \cdot \cdot \cdot k_d^{m_l}} \, .\label{eq:MZV_def}
\ee

Let us start by noting that if we have a space of (non-constant) functions at weight $p$, with basis elements
\be\label{eq:FunctionBasis}
F^{(p)}_{i_p}\,,\qquad i_p=1,2,\ldots, d_p\,,
\ee
then every weight-$m$ constant we add to this space as an independent basis element, will increase its dimension by
\be\label{eq:AddConstDim}
d_p\to d_p+d_{p-m}\,,
\ee
for $p\ge m$, with $d_0=1$. In other words, the addition of an independent constant, for example $\zeta_m$, also automatically implies the inclusion of a whole tower of functions,
\be
\zeta_m\, F^{(p-m)}_{i_{p-m}}\,,\qquad i_n=1,2,\ldots, d_{p-m}\,,
\ee
at higher weight. In our recursive construction of the function space, to date there exists no clean separation of this tower from the rest of the functions. The conclusion is thus that one should try to include as few of these independent constants in the function space as possible.

On the other hand, it is important to note that the promotion of all $n$-gon symbols of tables \ref{tab:HexSymbolDim} and \ref{tab:HepSymbolDim} to functions \emph{requires} the inclusion of certain independent constants. Let us now flesh out how this comes about, when carrying out this promotion by additionally imposing branch cut conditions such as eq.~\eqref{eq:hexBranchCuts} at each weight, as we have already mentioned. 

Concretely, suppose we have already constructed the basis of non-constant functions \eqref{eq:FunctionBasis} at weight $p$. Attaching another letter and solving the integrability conditions then gives a $\{p,1\}$ coproduct representation of the weight-$(p+1)$ space, and additionally evaluating them as indicated in the left-hand side of eq.~\eqref{eq:hexBranchCuts} yields a linear combination of the weight-$p$ functions in the soft limit where we impose our branch cut condition. This linear combination can be at most equal to a constant in the limit, since it is essentially equal to the discontinuity of the weight-$(p+1)$ function around the potential branch points where the letters, excluding those that can appear as first entries, vanish in the limit. Thanks to the function-level analog of eq.~\eqref{symbolDisc}, this discontinuity can indeed be proven to be a constant.

Sometimes it is possible to absorb this constant in a redefinition of the functions. For example, for $n=6$ and $p=2$ one may choose the basis \eqref{eq:FunctionBasis} as
\be
\biggl\{
\text{Li}_2\left(1-\frac{1}{u_i}\right),\ \log^2a_i \biggr\} \,,
\qquad i=1,2,3.
\label{eq:wt2Steinmann}
\ee
Then, we consider the subspace of these functions, spanned by the 
coproducts contributing to the branch cut condition~\eqref{eq:hexBranchCuts} of the functions at one weight higher. Furthermore, we restrict to the space of constant functions within this subspace. 

For our example, it turns out that three of the functions~\eqref{eq:wt2Steinmann} appear as $m_1,y_2$ or $y_3$ coproducts at weight $p+1=3$, and they are all constants in the corresponding soft limit~\eqref{eq:soft1},
\be\label{eq:soft1SpaceW2}
\biggl\{\text{Li}_2\left(1-\tfrac{1}{u_1}\right),\,\log^2a_2-\log^2a_3,\,4\text{Li}_2\left(1-\tfrac{1}{u_2}\right)+4\text{Li}_2\left(1-\tfrac{1}{u_3}\right)+\log^2a_1+\log^2a_2\,\biggr\}\to\{0,0,-8\zeta_2\} \,.
\ee
If we shift the functions \eqref{eq:wt2Steinmann} by a constant $c_i$ and $c'_i$, respectively then the right-hand side of eq.~\eqref{eq:soft1SpaceW2} becomes
\be\label{eq:soft1SpaceW2Eq}
\{c_1,c'_2-c'_3,4c_2+4c_3+c'_1+c'_2-8\zeta_2\} \,.
\ee
Any solution for $c_i, c'_i$, such that the above combinations are equal to zero, will therefore imply that all constant functions in the soft limit we are considering will vanish. The corresponding equations for the two cyclically related soft limits may be obtained from eq.~\eqref{eq:soft1SpaceW2Eq} by letting $c_i\to c_{i+1},\, c_4\equiv c_1$, similarly for the primed constants, and in this example it follows immediately that all three sets of equations have a unique solution $c_i=0$, $c'_i=4\zeta_2$. In other words, if we shift
\be\label{eq:lna_shift}
\log^2a_i\to \log^2a_i+4\zeta_2\,,
\ee
in our weight-2 basis \eqref{eq:wt2Steinmann}, then in the soft limit~\eqref{eq:soft1} and its cyclic permutations, all of its subspaces contributing to three weight-3 branch cut conditions of the type \eqref{eq:hexBranchCuts} will yield a space of  constant functions that shrinks to zero, and so these conditions will be automatically satisfied. Therefore we do not need to add $\zeta_2$ as an independent basis element at this weight.

While it is always possible to set the space of all constant functions in one limit to zero,\footnote{This is equivalent to choosing the base point of integration for our multiple polylogarithms to lie within this limit, and justifies why the MZVs we introduced in eq.~\eqref{eq:MZV_def} is sufficient for our discussion, as follows: The three soft limits are connected by collinear limits, where the space of functions reduce to so-called harmonic polylogarithms \cite{Remiddi:1999ew}. These in turn are known to evaluate to MZVs at their endpoints.} this is generally not the case for multiple limits. That is, at higher weight one cannot in general solve the analog of \eqref{eq:soft1SpaceW2Eq} and its cyclic images simultaneously. At best, we can solve a subset of these equations, whose rank equals the dimension of our basis of non-constant functions $d_p$. Then, the independent linear combinations of constants, analogous to eq.~\eqref{eq:soft1SpaceW2}, that the latter functions will reduce to, we will also need to include as elements in our basis at weight $p$. This will in turn guarantee that the branch cut conditions~\eqref{eq:hexBranchCuts} of all the weight-$(p+1)$ non-constant functions will have a solution, and that we can therefore maintain them in our space.

\renewcommand{\arraystretch}{1.25}
\begin{table}[!t]
\centering
\begin{tabular}[t]{l c c c c c c c c c c c c c c c}
\hline\hline
weight $n$
& 0 & 1 & 2 & 3 & 4 &  5 &  6 &  7 &  8 &  9 & 10 & 11 & 12 & 13 & 14
\\\hline\hline
$L=1$
& \green{1} & \green{3} & 4 &  &  &  &  &  &  &  &  &  &  &  & 
\\\hline
$L=2$
& \green{1} & \green{3} & \green{6} & 10 & 6 &  &  &  &  &  &  &  &  &  & 
\\\hline
$L=3$
& \green{1} & \green{3} & \green{6} & \green{13} & 24 & 15 & 6 &  &
&  &  &  &  &  & 
\\\hline
$L=4$
& \green{1} & \green{3} & \green{6} & \green{13} & \green{27} & 53
& 50 & 24 & 6 &  &  &  &  &  & 
\\\hline
$L=5$
& \green{1} & \green{3} & \green{6} & \green{13} & \green{27} & \green{54}
& 102 & 118 & 70 & 24 & 6 &  &  &  & 
\\\hline
$L=6$
& \green{1} & \green{3} & \green{6} & \green{13} & \green{27} & \green{54}
& \green{105} & 199 & 269 & 181 & 78 & 24 & 6 &  & 
\\\hline
$L=7+$
& \green{1} & \green{3} & \green{6} & \green{13} & \green{27} & \green{54}
& \green{105} & \green{200} & 338 & 331 & 210 & 85 & 27 & 6 & 1
\\\hline\hline
\end{tabular}
\caption{The number of independent $\{n,1,1,\ldots,1\}$ coproducts
of the MHV and NMHV amplitudes through $L=6$ loops. A green number
denotes saturation. The final line gives the number
using {\it all} known loop orders together, including 7 loop MHV.}
\label{tab:MHVNMHVdim}
\end{table}

Finally, let us emphasize that for a given amplitude normalization there exists an \emph{absolutely minimal} space of functions that contains the $L$-loop amplitude: This is spanned not only by the transcendental functions appearing in amplitude at weight $2L$, but also all of their $\{2L-k,1,\ldots,1\}$ coproducts. In other words, we also include the span of all the weight-$(2L-1)$ functions in the derivative of the amplitude, then compute all of their derivatives and construct the span again, and repeat $k$ times. The dimensions of these spaces, based on the existing data and for the cosmic normalization, which will be justified in the next subsection, are depicted in table \ref{tab:MHVNMHVdim}. We observe that the dimension of the weight-$(2L-k)$ function space
generated in this fashion increases with $k$ until it saturates at $k=L$. Conversely, if we drop any of the weight-$k$ functions of the saturated space when building our space bottom-up, by increasing the weight by one at each step, we are guaranteed that at weight $2k$, this space will not be large enough to contain the amplitude.

So for a given normalization, this top-down exploration of the span of the amplitude and its iterated coproducts is useful because it exposes the set of independent constants we need to include in our space, which may even be smaller than what is required to promote every symbol into a function. More importantly however, this span changes in size as we change normalization. It can can therefore guide the choice of optimal normalization, in conjunction with the coaction principle, as we will now explain.

\subsection{The coaction principle}

In section \ref{sec:MPLs}, we mentioned that hexagon and heptagon function spaces belong to the broader class of multiple polylogarithms, which admit a natural coaction operation $\Delta$ on them, and particularly focused on its $\Delta_{m,1,\ldots,1}$ components. When restricting to classical polylogarithms, the full coaction is also simple to write down,
\be\label{eq:Delta_Li_n}
\Delta(\textrm{Li}_{n}(z)) = 1\otimes \textrm{Li}_n(z) + \sum_{k=0}^{n-1}\textrm{Li}_{n-k}(z)\otimes \frac{\log^kz}{k!}\,,
\ee
whereas for generic MPLs it may be found in the original literature \cite{Gonch3}, or for example in the review~\cite{Duhr:2014woa}. At the practical level, it has also been implemented in the \texttt{Mathematica} package \texttt{PolyLogTools}~\cite{Duhr:2019tlz}, see also the \texttt{HyperlogProcedures}~\cite{HyperlogProcedures} package in \texttt{Maple}, as well as the latter's native MPL functionality~\cite{Frellesvig:2018lmm}.

What is more, the coaction naturally also extends to the transcendental constants MPLs evaluate to, in particular it is consistent to define~\cite{Brown:2011ik,Duhr:2012fh}
\begin{equation}\label{eq:Delta_pi}
\Delta(i\pi)=(i\pi)\otimes 1\,,
\end{equation}
whereas from eqs.~\eqref{eq:ZetaLi1} and \eqref{eq:Delta_Li_n} it follows that
\be\label{eq:Delta_zeta}
\Delta(\zeta_{2m+1}) = 1\otimes \zeta_{2m+1} + \zeta_{2m+1} \otimes 1\,.
\ee
Since the coaction is compatible with multiplication,
\be
\Delta(a b)=\Delta(a)\Delta(b)\,,
\ee
with the latter defined component-wise on the right hand side, the last two equations, for example, imply
\be\label{DeltaZetaSquare}
\quad \Delta(\zeta_{2m+1}^2)=(\zeta_{2m+1}^2)\otimes 1+2 \zeta_{2m+1}\otimes \zeta_{2m+1}+ 1\otimes (\zeta_{2m+1}^2)\,.
\ee

While the coaction is valid by construction on the space of all multiple polylogarithms $\cG$, a rather nontrivial closure has been observed on certain subspaces $\cH\subset \cG$,
\be
\boxed{
\Delta \cH \subset \cH \otimes \big[\cG\mod\,(i\pi)\big].}
\label{eq:coaction_principle}
\ee
The factor on the right may in fact carve out only a subspace of $\cG$, but the crucial aspect of the above statement lies in the left factor: For any element in $\cH$, the left factor of the coaction must also lie in $\cH$. This is the \emph{(cosmic Galois) coaction principle}, which can be shown to be equivalent to the action of a symmetry group on $\cH$, analogous to the action of the Galois group on solutions to polynomial equations~\cite{Brown:2015fyf}. It has hence been coined the ``cosmic Galois group'' by mathematician Pierre Cartier \cite{Cartier2001}.

If the coaction principle \eqref{eq:coaction_principle} holds, it implies powerful constraints that have the form of exclusion principles, or ``superselection rules'', in Cartier's words. For example, assume that at a given instance we have knowledge of $\cH$ up to weight 3, and we observe that it does not contain the constant $\zeta_3$. Then, since $\zeta_3$ appears on the left factor of the coaction on $\zeta_3^2$, eq.~\eqref{DeltaZetaSquare}, the coaction principle predicts that $\zeta_3^2$ will \emph{not} appear in $\cH$ at weight 6.

Indeed, constraints of this type have been observed to hold in the space spanned by the coefficient of the logarithmic divergence of certain vacuum graphs with $2L$ edges at $L$ loops~\cite{Schnetz:2013hqa,Panzer:2016snt}, as well as by the polylogarithmic part of the anomalous magnetic moment of the electron through four loops~\cite{Schnetz:2017bko}. They have also been observed in tree-level scattering amplitudes in string theory, connecting different orders in the expansion with respect to the inverse string tension $\alpha'$~\cite{Schlotterer:2012ny}. Finally, the coaction principle has been proved for certain Feynman integrals in a motivic setting~\cite{Brown:2015fyf}.

These occurrences of the coaction principle offer ample motivation to explore whether it also holds when $\cH$ is the function space containing the $\cN=4$ SYM amplitudes and their derivatives. 	Of course, part of its content holds automatically due to the fashion in which we construct this space, where functions at weight $m$ are promoted to $\Delta_{m,1}$ components at one weight higher. The novel import of~\eqn{eq:coaction_principle} resides in the fact that, as shown in eqs.~\eqref{eq:Delta_pi}-\eqref{DeltaZetaSquare}, transcendental constants also exhibit structure under the coaction map, even though they are in the kernel of $\Delta_{m,1}$.

Since non-constant functions are in any case captured by $\Delta_{m,1}$ or more generally $\Delta_{m,1,\ldots,1}$, we will therefore search for nontrivial manifestations of the coaction principle at certain kinematic points, where the function spaces containing our amplitudes evaluate to constants. Focusing on the amplitude with $n=6$ particles, in particular we will examine the point $a_i=u_i=1$ or $(1,1,1)$, which is arguably the most natural place to look, since all functions remain finite there, and it is also invariant under dihedral transformations. In table \ref{table:H_zeta_space_constants}, we display what the saturated space of  amplitudes and their derivatives, in the sense we discussed in the previous section and in table \ref{tab:MHVNMHVdim} more specifically, evaluates to at (1,1,1) up to weight 5\footnote{Up to this weight, the BDS-like and cosmic normalization don't differ, so the data presented in table \ref{table:H_zeta_space_constants} holds for both.}. In addition we display the constants we have to include as independent basis elements in our function space, as we described in the previous subsection. Note that while the latter constants must necessarily also appear at $(1,1,1)$, the converse is not true.
\begin{table}
\begin{center}
\begin{tabular}{  r  c  c  c  } 
    \hline\hline
{\footnotesize Weight} & {\footnotesize \ \ Multiple Zeta Values \ \ } & {\footnotesize Appear at $(1,1,1)$} & {\footnotesize Independent Constants} \\ 
\hline
0 & 1 & 1 & 1 \\ 
\hline
1 & $-$ & $-$ & $-$ \\ 
\hline
2 & $\zeta_2$ & $\zeta_2$ & $-$ \\ 
\hline
3 & $\zeta_3$ & $-$ & $-$ \\ 
\hline
4 & $\zeta_4$ & $\zeta_4$ & $\zeta_4$ \\ 
\hline
5 & $\zeta_5$, $\zeta_2 \zeta_3$ & $5 \zeta_5 - 2 \zeta_2 \zeta_3$ & $-$ \\ 
\hline\hline
\end{tabular} 
\caption{For weight $m\le 5$, we display first the complete set of MZVs, followed
by the linear combinations that appear in $\mathcal{H}_{6,m}$
when the functions are evaluated at $(1,1,1)$, followed
by the independent constants contained in this space.}
\label{table:H_zeta_space_constants}
\end{center} 
\end{table}

A first observation from table \ref{table:H_zeta_space_constants} is that $\zeta_2$ is not needed as an independent constant in the space of amplitudes and their derivatives. This was also the case when the space of Steinmann hexagon symbols was minimally promoted to functions in the previous subsection, and in fact the two spaces can be shown to coincide up to the weight shown in the table. In particular, notice that the difference between the dimension the symbols of table \ref{tab:HexSymbolDim} and of the saturated functions of table \ref{tab:MHVNMHVdim}, is precisely accounted for by the addition of $\zeta_4$ as a basis element, see also eq.~\eqref{eq:AddConstDim}.

Secondly, notice that if a coaction principle were to hold in the space of the BDS-like normalized amplitude at the kinematic point (1,1,1), then the absence of $\zeta_3$ would also imply the absence of $\zeta_3^2$, as we explained earlier in this section. Unfortunately, this is not the case, as the 3-loop MHV amplitude evaluates to
\be
\cE_{6,0}^{(3)}(1,1,1) = \frac{413}{3} \, \zeta_6 + 8 (\zeta_3)^2 \,.\\
\ee
The presence of this $\zeta_3^2$ is also at odds with our first observation. Namely the latter constant has to be added as an independent basis element on top of what is minimally needed to as to promote all hexagon symbols to functions.

Therefore, the coaction principle strongly motivates absorbing the factor proportional to $\zeta_3^2$ in a redefinition of the normalization,
\be
\begin{aligned}
\cE_{6,0}^{(3)}(1,1,1)& \equiv \left(\rho \cE'(1,1,1)\right)^{(3)}=\cE'^{(3)}(1,1,1)+\rho^{(3)}\,\,\,\text{with}\,\,\,\rho=1+ 8 (\zeta_3)^2g^6+\cO(g^8)\,.
\label{old3loops111}
\end{aligned}
\ee
On the right-hand side, we recognize the first correction to the factor \eqref{rho} governing the cosmic normalization \eqref{eq:CosmicNorm}. This is how the coaction principle allows us to determine this normalization also at higher loops, when combined with a few more reasonable assumptions~\cite{Caron-Huot:2019vjl}, in parallel with the amplitude\footnote{As we have already mentioned, more recently a finite-coupling conjecture for $\rho$ has appeared~\cite{Basso:2020xts}.}.

We wish to stress that changing from BDS-like to cosmic normalization is not merely a reorganization of the same information, similarly to how changing from BDS to BDS-like was not. Had we not carried out the former change, we would have had to include $\zeta_3^2$ (and further constants appearing in $\rho$ beyond three loops) as an independent constant when building the function space from the bottom up, which would also induce a whole tower of extra functions at higher weight, as per the discussion around eq.~\eqref{eq:AddConstDim}. So the cosmic normalization \emph{reduces the size} of the space we need to need to bootstrap, and thus facilitates the identification of the amplitude within this space.

For the sake of simplicity so far we only talked about the MHV amplitude, but it's worth emphasizing that normalization by the same factor $\rho$ simultaneously renders the analogous function entering the description of the NMHV amplitude, denoted as $E'^{(L)}$ (with the prime sometimes dropped), compatible with the coaction principle as well, at least up to the loop order so far computed. The fact that the same factor is required for both helicity configurations points towards its ubiquity; and assuming it holds at higher loops, it also predicts that the difference $\cE'^{(L)}(1,1,1)-E'^{(L)}(1,1,1)$ respects the coaction principle even when computed by truncating $\rho$ at one lower loop order.

Having analyzed the role of transcendental constants, and of the coaction principle in most efficiently absorbing them into the normalization of the amplitude, we can now spell out explicitly our complete bottom-up definition of the space of functions $\cH_{6,m}$ relevant for bootstrapping six-particle amplitudes. The space $\cH_{6,m}$ inherits the properties and definitions already discussed in sections \ref{sec:HexHepFuns} and \ref{sec:SteinClus}, namely it consists of multiple polylogarithms made of the 9-letter symbol alphabet of eq.~\eqref{eq:hexletters} plus cyclic, obeying the first entry condition, eq.~\eqref{eq:firstentries}, the extended Steinmann relations \eqref{eq:FabSteinw2}, as well as integrability. In addition, we require that functions in this space satisfy branch cut conditions such as eq.~\eqref{eq:hexBranchCuts}, and only include even zeta values except $\zeta_2$, namely
\be
\zeta_4 \,,\ \zeta_6 \,,\ \zeta_8 \,,\ \zeta_{10} \,,\ \zeta_{12} \,,\ \ldots.
\label{eq:indepzetas}
\ee
as independent constants. We have constructed $\cH_{6,m}$ up to weight $m=12$ at function level and up to $m=13$ at symbol level, as well as $m=14$ when restricting the final symbol entry to the subset known to be relevant for MHV amplitudes. We have confirmed that it respects the coaction principle not only at point $(1,1,1)$, but also in several other points and lines in the space of kinematics. We form an ansatz for the amplitude in the BDS-like normalization, where boundary data for the latter is available, expressing it in terms of a basis in $\cH_{6,m}$ with undetermined coefficients, times the normalization conversion factor $\rho$, which may also contain certain undetermined coefficients. By comparing with the boundary data, we fix all undetermined coefficients, and thus both the cosmically normalized amplitude and $\rho$, as we move on to discuss next.

\section{Results}
In the previous two sections, we exposed the conceptual developments underlying the most recent refinement of the amplitude bootstrap: On the one hand the extended Steinmann relations, or equivalently cluster adjacency, which maximally simplify the spaces of symbols in which the amplitudes live; and on the other hand the coaction principle, which similarly guides the elimination of any redundancies in the corresponding function-level spaces. Let us now present two concrete applications of these developments, the computation of the six-particle (N)MHV amplitude to (six) seven loops, and the symbol of the seven-particle NMHV amplitude to four loops.

\subsection{The Six-Particle Amplitude at Six and Seven Loops}
\renewcommand{\arraystretch}{1.25}
\begin{table}[!t]
\centering
\def\sp{@{{\,}}}
\begin{tabular}[t]{l\sp\sp c\sp\sp c\sp\sp c\sp\sp c\sp\sp c\sp\sp c\sp\sp c\sp\sp c}
\hline
Constraint                      & $L=1$\, & $L=2$\, & $L=3$\, & $L=4$\, & $L=5$\, & $L=6$
\\\hline
1. $\cH_6$ & 6 & 27 & 105 & 372 & 1214 & 3692?\\\hline\hline
2. Symmetry       & (2,4) & (7,16) & (22,56) &  (66,190) & (197,602)& (567,1795?)\\\hline
3. Final-entry       & (1,1) & (4,3) & (11,6) &  (30,16) & (85,39) & (236,102)\\\hline
4. Collinear       & (0,0) & (0,0) & $(0^*,0^*)$ &  $(0^*,2^*)$ & $(1^{*3},5^{*3})$ & $(6^{*2},17^{*2})$ \\\hline
5. LL MRK       & (0,0) & (0,0) & (0,0) &  (0,0) & $(0^*,0^*)$ & $(1^{*2}$,$2^{*2})$ \\\hline
6. NLL MRK       & (0,0) & (0,0) & (0,0) &  (0,0) & $(0^*,0^*)$ & $(1^*,0^{*2})$ \\\hline
7. NNLL MRK       & (0,0) & (0,0) & (0,0) &  (0,0) & (0,0) & $(1,0^*)$ \\\hline
8. N$^3$LL MRK       & (0,0) & (0,0) & (0,0) &  (0,0) & (0,0) & (1,0) \\\hline
9. Full MRK       & (0,0) & (0,0) & (0,0) &  (0,0) & (0,0) & (1,0) \\\hline
10. $T^1$ OPE       & (0,0) & (0,0) & (0,0) &  (0,0) & (0,0) & (1,0) \\\hline
11. $T^2$ OPE       & (0,0) & (0,0) & (0,0) &  (0,0) & (0,0) & (0,0) \\\hline
\end{tabular}
\caption{Remaining parameters in the ans\"{a}tze for
the (MHV, NMHV) amplitude after each constraint is applied,
at each loop order. The superscript ``$*$'' (``$*n$'') denotes an additional
ambiguity ($n$ ambiguities) which arises only due to lack of knowledge
of the cosmic normalization constant $\rho$ at the given stage. The ``$?$'' indicates an ambiguity about
the number of weight 12 odd functions that are ``dropouts''; they
are allowed at symbol level but not function level. The seven-loop MHV amplitude was constrained in a somewhat different order. As the parameter counts are not directly comparable it is omitted from the table.}
\label{tab:full}
\end{table}

In table \ref{tab:full} we detail the sequence of steps taken in order to find a unique solution for our ansatz for the $L$-loop $n=6$ (MHV, NMHV) amplitude \cite{Caron-Huot:2019vjl}, along with the number of undetermined parameters at any given step. Line 1 is equal to the dimension of the bottom-up space $\cH_{6,2L}$, whose construction we described in sections \ref{sec:HexHepFuns} through \ref{sec:Coaction}. Line 2 amounts to imposing the discrete dihedral and parity symmetries of the amplitude, which we reviewed in section \ref{sec:SymKin}. The third line refers to constraints on the final entry of the symbol, arising from dual superconformal symmetry in the form of the $\bar Q$-equation, as we briefly mentioned in the introduction, and we will expand on particularly for $n=7$ in the next subsection.

All remaining lines refer to particular kinematic limits where we may obtain independent information on the behavior of the amplitude, starting with the strict collinear limit, whose divergences are completely captured by the BDS ansatz. Then follows the high energy or multi-Regge kinematics (MRK), which is a very rich subject owing to the development of an effective description of the latter by Balitsky, Fadin, Lipatov and Kuraev in QCD, that was then extended to planar $\cN=4$ SYM~\cite{Bartels:2008sc,Fadin:2011we,Bartels:2011ge,Lipatov:2012gk,Dixon:2012yy,Bartels:2013jna,Dixon:2014iba,Drummond:2015jea,DelDuca:2016lad,DelDuca:2018hrv}. Because of dual conformal symmetry, it is equivalent to the soft limit \eqref{eq:soft1}, and one needs to analytically continue away from the Euclidean region in order to obtain a nontrivial result. As can be seen from the aforementioned limit, it is natural to organize perturbative expansion of the amplitude also with respect to the order of the divergent logarithm, $\log^{L-p-1} a_1$, denoted as the (next-to)$^p$-leading-logarithmic (N$^p$LL) approximation. A remarkable consequence of the integrability of the theory is that this double expansion can be computed at any loop order and logarithmic approximation, not only for $n=6$~\cite{Basso:2014pla}, but also at arbitrary multiplicity~\cite{DelDuca:2019tur}.

Finally, the table contains the expansion around the collinear limit, arranged in dofferent powers of a particular cross ratio $T\to 0$, which is governed by the Pentagon OPE \cite{Alday:2010ku,Basso:2013vsa,Basso:2013aha,Basso:2014koa,Basso:2014nra,Belitsky:2014sla,Belitsky:2014lta,Basso:2014hfa,Basso:2015rta,Basso:2015uxa,Belitsky:2016vyq} also mentioned in the introduction. That is, while the resummation of the entire OPE is a challenging endeavor, individual terms in this expansion can be straightforwardly evaluated using the methods of \cite{Papathanasiou:2013uoa,Papathanasiou:2014yva}. As we move down each column of the table, the complete determination of the amplitude for both helicity configurations at the given loop order occurs with the first $(0,0)$ table entry we encounter. Any entries below the latter then provide consistency checks of our unique solution.

The table clearly shows the difference between six loop MHV and all other cases shown, in that one parameter survives all the way through the multi-Regge
limit and  $\cO(T^1)$ OPE constraints.  The same is true
for the  seven-loop MHV amplitude, whose parameter counts we have not included in the table, since it was computed somewhat differently.

\begin{figure}
\begin{center}
\includegraphics[width=5.5in]{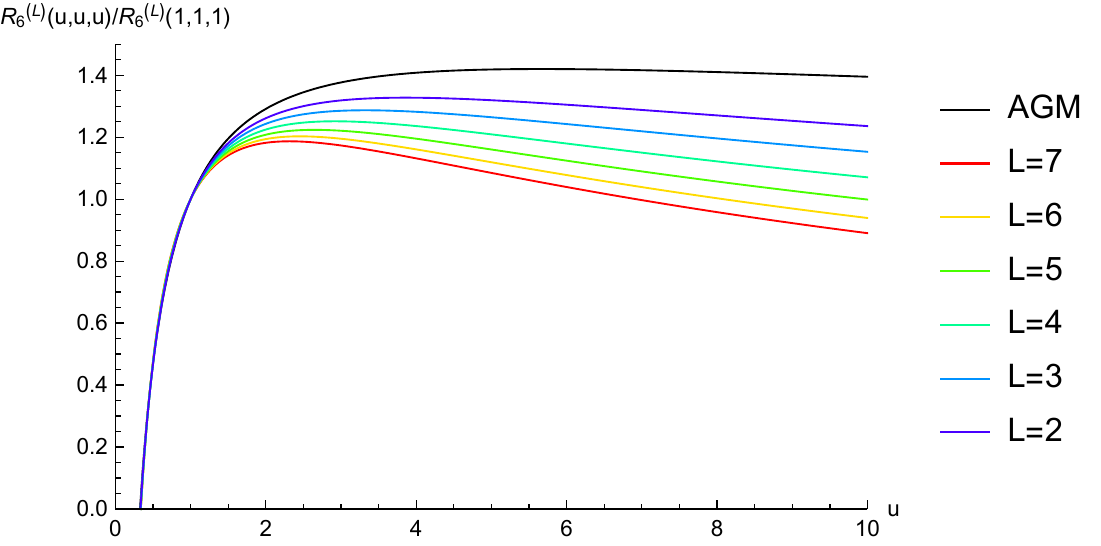}
\end{center}
\caption{Normalized perturbative coefficients of the remainder function, $R_6^{(L)}(u,u,u)/R_6^{(L)}(1,1,1)$, for $L=2$ to 7, plotted along with the strong-coupling result of AGM. The curves all have a remarkably similar shape for $u\lesssim 1$.}
\label{fig:r6uuu}
\end{figure}

With the new six- and seven-particle amplitudes at hand, we may proceed to 
analyze their behavior at various interesting kinematic subspaces and points, either numerically or analytically. As an illustration, in figure \ref{fig:r6uuu} we plot the remainder function \eqref{eq:R6} on the line $a_i=1/u$ or $u_i=u$, normalized by its value at $u=1$, across loop orders, also including the strong coupling prediction of Alday, Gaiotto and Maldacena (AGM)~\cite{Alday:2009dv}.  Once normalized in this way, the functions are almost indistinguishable for $u<1$, and maintain quite similar shapes also for $u>1$.

On the analytic side, one rather interesting observation was that at the origin in the space of cross ratios \eqref{eq:udef}, namely the limit where all three $u_i\to 0$, the remainder function (i.e. logarithm of the BDS-normalized MHV amplitude) behaves as
\be
{\cal R}_{6} = c_{1} P_{1} + c_{2} P_{2} + c_0 + {\cal O}(u_i) \,,
\ee
where $P_i$ denotes the two symmetric quadratic polynomials in $\log{u_i}$,
\be
P_{1} = P_{2} + \sum_{i=1}^{3} \log^2{u_{i}}\, ,
\quad P_{2} = \sum_{i=1}^3 \log{u_{i}}\log{u_{i+1}}\,,
\ee
and $c_i^{(L)}$ are constants, in particular combinations of zeta values.
For example, their values at six and seven loops are:
\bea
c_1^{(6)} &=& \frac{2033119}{160} \, \zeta_{10} - 159 \, \zeta_4 (\zeta_3)^2
   - 240 \, \zeta_2 \zeta_3 \zeta_5 - 420 \, \zeta_3 \zeta_7
   - 204 \, (\zeta_5)^2 \,,  \label{c_1_6}\\
c_1^{(7)} &=& - \frac{8404209697}{44224} \, \zeta_{12} + 1620 \, \zeta_6 (\zeta_3)^2 + 3252 \, \zeta_4 \zeta_3 \zeta_5 + 2520 \, \zeta_2 \zeta_3 \zeta_7 \nonumber\\
&&\null\hskip0.0cm + 1224 \, \zeta_2 (\zeta_5)^2 + 4704 \, \zeta_3 \zeta_9 + 4368 \, \zeta_5 \zeta_7 + 20 \, (\zeta_3)^4 \,,  \label{c_1_7}
\eea
\bea
c_2^{(6)} &=& \frac{2532489}{80} \, \zeta_{10} + 126 \, \zeta_4 (\zeta_3)^2
   + 160 \, \zeta_2 \zeta_3 \zeta_5 \,,  \label{c_2_6}\\
c_2^{(7)} &=& - \frac{9382873343}{22112} \, \zeta_{12} - 1360 \, \zeta_6 (\zeta_3)^2 - 2568 \, \zeta_4 \zeta_3 \zeta_5 - 1680 \, \zeta_2 \zeta_3 \zeta_7 \nonumber\\
&&\null\hskip0.0cm - 816 \, \zeta_2 (\zeta_5)^2 - 8 \, (\zeta_3)^4 \,,  \label{c_2_7}
\eea
\bea
c_0^{(6)} &=& \frac{54491355251}{265344} \, \zeta_{12}
    - \frac{5741}{4} \, \zeta_6 (\zeta_3)^2 - 3620 \, \zeta_4 \zeta_3 \zeta_5
    - 3780 \, \zeta_2 \zeta_3 \zeta_7 \nonumber\\
&&\null\hskip0.0cm
    - 1836 \, \zeta_2 (\zeta_5)^2
    - 4704 \, \zeta_3 \zeta_9 - 5208 \, \zeta_5 \zeta_7
    - 14 \, (\zeta_3)^4 \,, \label{c_0_6}\\
c_0^{(7)} &=& - \frac{3768411721}{1280} \, \zeta_{14} + \frac{52815}{4} \, \zeta_8 (\zeta_3)^2 + 31187 \, \zeta_6 \zeta_3 \zeta_5 + 38850 \, \zeta_4 \zeta_3 \zeta_7 \nonumber\\
&&\null\hskip0.0cm + 18750 \, \zeta_4 (\zeta_5)^2 + 42336 \zeta_2 \zeta_3 \zeta_9 + 39312 \, \zeta_2 \zeta_5 \zeta_7 + 156 \, \zeta_2 (\zeta_3)^4\nonumber\\
&&\null\hskip0.0cm
+ 55440 \, \zeta_3 \zeta_{11} + 61824 \, \zeta_5 \zeta_9 + 31560 \, (\zeta_7)^2 + 560 \, (\zeta_3)^3 \zeta_5\,. \label{c_0_7}
\eea
Apart from the relative number-theoretic simplicity of the coefficients, the important aspect to note is that the degree of the divergent logarithms is always two, independent of the loop order, and that no linear term is present. Based on these observations, and harnessing the power of the Pentagon OPE, the authors of~\cite{Basso:2020xts} have very recently conjectured the form of the coefficients $c_i$ at finite coupling.

\

\subsection{The Symbol of the Seven-Particle NMHV Amplitude at Four Loops}

In section \ref{sec:SteinClus}, we mentioned that the extended Steinmann relations are equivalent to the property of cluster adjacency, when considering their effect on the transcendental functions appearing in $\cN=4$ SYM amplitudes and integrals contributing to them. In fact, there exists one aspect in which cluster adjacency is somewhat stronger: As it has become apparent more recently \cite{Drummond:2018dfd}, it also constrains any potential rational factors in the amplitude\footnote{For more recent work in this direction see also \cite{Mago:2019waa,Lippstreu:2019vug,Lukowski:2019sxw}.}, as well as relates the rational and transcendental parts. This becomes particularly relevant for non-MHV amplitudes, which indeed have such nontrivial rational parts.

So in this subsection, we will precisely describe how this cluster adjacency property can be exploited in order to construct the symbol of the seven-particle NMHV amplitude $\cE_{7,1}$ at four loops~\cite{Drummond:2018caf}.  To this end, we will need to first briefly recall the general structure of $\cE_{n,1}$.

The rational part of superamplitudes in $\cN=4$ SYM theory respects both superconformal and dual superconformal symmetry, in other words it is Yangian invariant~\cite{Drummond:2010qh}. As such, it is best described by generalizing momentum twistors to supertwistors,
\be
Z_i\to \mathcal{Z}_i = (Z_i \, | \, \chi_i)\,,
\ee
where the fermionic components $\chi_i$ are related to fermionic dual coordinates, very similarly to how the bosonic twistors $Z_i$ are related to the bosonic dual coordinates $x_i$. Particularly for $\cE_{n,1}$, the building blocks of its rational part are known as $R$-invariants, and for momentum supertwistors $\cZ_a, \ldots, \cZ_e$, they may be defined as~\cite{Drummond:2008vq,Drummond:2008bq,ArkaniHamed:2010kv}
\be\label{fivebrak}
[abcde] = \frac{\delta^{0|4} \big( \chi_a \langle b c d e \rangle + \text{cyclic} \big)}{\langle a b c d \rangle \langle b c d e \rangle \langle c d e a \rangle \langle d e a b \rangle \langle e a b c \rangle} \,,
\ee
where the bosonic four-bracket has already been defined in \eqref{fourbrak}, and the fermionic delta function is given by $\delta^{0|4}(\xi)=\xi^1 \xi^2 \xi^3 \xi^4$ for the different $SU(4)$ components of $\xi$. Thus $\chi_i$ serves the purpose of packaging the different bosonic components of the superamplitude together, similarly to the Gra\ss mann variable $\eta$ in \eqref{eq:PhiSuperfield}, and it can in fact be expressed in terms of the latter.

Not all $R$-invariants are independent, as they obey identities of the form~\cite{Drummond:2008vq,Mason:2009qx}
\be\label{sixZidentity}
[abcde] - [bcdef] + [cdefa] - [defab] + [efabc] - [fabcd] = 0\,,
\ee
for any set of six supertwistors. The identities are in turn not all independent, leaving \cite{Elvang:2010xn}
\be
\#\,\,\text{linearly independent}\,\,n\text{-particle}\,\,R\text{-invariants}=\binom{n-1}{4}\,,
\ee
namely 5 and 15 independent $R$-invariants relevant for  NMHV scattering amplitudes with $n=6$ and $n=7$ external legs, respectively. 

Focusing on $n=7$ from now on, we may simplify the notation by denoting the $R$-invariants by the two points not contained in them,
\begin{equation}
  (12) = [34567]\,,
  \qquad
  (13) = [24567]\,,
  \qquad
  (14) = [23567]\,,
\end{equation}
and similarly for their cyclic copies. By convention, we will assume that the right-hand side is always ordered, so that ordering on the left-hand side doesn't matter.

With the help of the above definitions, the (non-manifestly cyclically symmetric) tree-level superamplitude may be written as~\cite{Drummond:2008bq}
\begin{equation}  
{\cal E}^{(0)}_{7,1} = (12) + (14) + (34) + (16) + (36) + (56)\,,\\
\end{equation}
and more generally at higher loops one may write
\begin{equation}
  \label{eq:ansatz}
  \mathcal{E}_{7,1}^{(L)} =
  e_{12}^{(L)} \, (12) + e_{13}^{(L)} \, (13) + e_{14}^{(L)} \, (14) + \text{cyclic},
\end{equation}
bearing in mind, however, that the quantities $e_{ij}$ are not uniquely defined, due to the identities \eqref{sixZidentity}. This ambiguity can be resolved by expressing all $R$-invariants in terms of a basis of 15 thereof, and a convenient choice of basis in the literature~\cite{CaronHuot:2011kk} consists of ${\cal E}^{(0)}_{7,1}$, as well as $(12)$, $(14)$, and their cyclic images. Then, eq.~\eqref{eq:ansatz} takes the form~\cite{Dixon:2016nkn}
\begin{equation}
  \label{eq:heptNMHVform}
  \mathcal{E}_{7,1}^{(L)} =
  E_0^{(L)}  {\cal E}^{(0)}_{7,\text{1}} + \bigl(  E_{12}^{(L)} \, (12) + E_{14}^{(L)} \, (14) + \text{cyclic} \bigr)\,,
\end{equation}
where now the coefficients of the $R$-invariants are well-defined symbols in $\cH_{7,2L}$, and are related to the objects $e_{ij}$ as
\begin{equation}\label{eq:Etoe}
  E^{(L)}_0   = \sum_{i=1}^{7} e^{(L)}_{i\,i+2},
  \qquad
  E^{(L)}_{14} = e^{(L)}_{14} - e^{(L)}_{16} - e^{(L)}_{46},
  \qquad
  E^{(L)}_{12} = e^{(L)}_{12}- e^{(L)}_{16} - e^{(L)}_{24} - e^{(L)}_{46}\, .\end{equation}
Next, let us proceed to explain how to compute these nontrivial building blocks of $\mathcal{E}_{7,1}^{(L)}$ at $L=4$. The first step in which cluster adjacency plays an instrumental role, is in the simplification of the $\bar Q$-equation \cite{CaronHuot:2011kk}, which governs the violation of dual superconformal symmetry at loop level, due to infrared divergences. In particular, this equation constrains the pairs of final symbol entries times  R-invariants  that can appear in the amplitude, to just a set of $147$ instead of the total $42\times 15=630$ naively allowed. For $\mathcal{E}_{7,1}^{(L)}$ these pairs have been presented in~\cite{Dixon:2016nkn}, however in a form involving several rather complicated combinations of such pairs.

Instead, by making use of the identities \eqref{sixZidentity}, these can be recast in the most elementary, manifestly cluster adjacent form of a single (final entry)$\otimes$($R$-invariant) pair \cite{Drummond:2018dfd}:
\begin{align}
  \hnsqbar{(12)} &= \{ a_{15}, a_{21}, a_{26}, a_{32}, a_{34}, a_{53}, a_{57}\}
                  \notag \\
  \hnsqbar{(13)} &= \{ a_{21}, a_{23}, a_{31}, a_{33}, a_{41}, a_{43}, a_{62}\}
                  \notag\\
  \hnsqbar{(14)} &= \{ a_{11}, a_{14}, a_{21}, a_{24}, a_{31}, a_{34}, a_{46}\}
                   \qquad  \text{\& cyclic}\,,
                   \label{eq:qbarca}
\end{align}
where the right-hand side of the first equation lists all final entries that can appear next to $(12)$ and so on. The notation on the left-hand side signifies that these lists are contained in the homogeneous neighbor sets of the $R$-invariants, in the sense that they are letters that can appear in a cluster also containing all the $\cA$-coordinates in the denominator of \eqref{fivebrak}. The subscript denotes that the $\bar Q$-equation is stronger, namely it only selects a subset of all independent cluster adjacent pairs.

The second manner with which cluster adjacency significantly facilitates the computation, is by automatically solving all the simplest equations for the double coproducts of all the functions $F \in \cH_{n,m}$, eq.~\eqref{DoubleCopMatrix}, in particular those that have the form
\be
F^{\overline{\phi}_\beta,\phi_\beta}=0\,,
\ee
where $\overline{\phi}_\beta$ does not belong to the neighbor set of ${\phi}_\beta$. This is achieved by forming an ansatz
\be
F=\sum_{\beta}\sum_i c_{i,\beta}F^{(i)}_{\text{hns}(\phi_\beta)}\otimes\phi_\beta\,,
\ee
where $F^{(i)}_{\text{hns}(\phi_\beta)}$ are elements of a basis of functions labeled by $i$ at one weight less, whose final entries are in the homogeneous neighbor set of the letter $\phi_\beta$,
\be
dF^{(i)}_{\text{hns}(\phi_\beta)}=\sum_{\phi_\alpha\in \text{hns}(\phi_\beta)} F^{(i),\phi_\alpha}d\log\phi_\alpha\,.
\ee
This is the very useful notion of a \emph{neighbor set function}, which reduces the size of the linear system we have to solve in order to construct $\cH_{n,m}$ at each weight.

All in all, our ansatz for the transcendental part of ${\cal E}^{(L)}_{7,1}$ with $L=4$ will have the form
\begin{equation}
  e_{ij}^{(L)} =
  \sum_{\phi_\alpha\, \in\, \hnsqbar{(ij)}}\,\,
   \sum_{k}
  \,\,
  c^{(ij)}_{k, \alpha}\,\, f^{(k)}_{\hns{\phi_\alpha}}  \otimes \phi_{\alpha}\,,
\end{equation}
where $f^{(k)}_{\hns{\phi_\alpha}}$ are weight-$(2L-1)$ neighbor set functions, and the final entries $\phi_\alpha$
are chosen from the set $\hnsqbar{(ij)}$ defined in equation
(\ref{eq:qbarca}). Note that we have started with the non-unique objects appearing in \eqref{eq:ansatz}, simply because the latter equation has a simple factorized form only with respect to the 21 distinct $R$-invariants, instead of the 15 independent ones. Once we have constructed the ansatz, however, one may switch to the well-defined components \eqref{eq:Etoe} at any stage.

Our initial ansatz, also incorporating certain reflection symmetries, contains 16,212 undetermined coefficients. Consecutively imposing integrability on $E_{14}$ and $E_{12}$ reduces this number to 8,444 and 56, respectively. Then, eliminating certain spurious poles of the $R$-invariants \cite{Korchemsky:2009hm,Dixon:2016nkn} brings this number down to 5. Finally, constraints from the strict collinear limit uniquely fix the amplitude. As a consistency check, we have compared the multi-Regge limit of our answer to existing predictions in the literature~\cite{DelDuca:2018hrv} up to next-to-leading logarithm. In addition, we have obtained new predictions up to (next-to)$^3$-leading-logarithmic accuracy.

\section{Conclusion}
The amplitude bootstrap has provided a new paradigm for tremendously improving perturbative quantum field-theoretic computations, by exploiting the analytic structure of the considered physical quantity. Developed first for the six- and then for the seven-particle amplitude in planar $\cN=4$ SYM theory, it succeeded not only in determining these amplitudes to an unprecedented loop order, but perhaps more importantly, to reveal new physical and mathematical properties, such as the extended Steinmann relations/cluster adjacency, or the coaction principle, with potential wider applicability. It will be very exciting to explore these properties, and expand the realm of the paradigm, in the setting of more realistic gauge theories. Indeed, some highly encouraging first applications of the bootstrap philosophy to QCD include the computation of the soft anomalous dimension at 3 loops~\cite{Almelid:2017qju}, as well in the construction of the space of functions/integrals describing massless five-particle scattering~\cite{Chicherin:2017dob}. And perhaps even more remarkably, as the final lines of this contribution were being written, the extended Steinmann relations were confirmed for the (planar) generalization of the aforementioned space, when one of the external legs is massive~\cite{Abreu:2020jxa}.

It is also natural to ask if the power of the bootstrap continues as the multiplicity increases, even in $\cN=4$ SYM. For some time, a conceptual obstacle was that for higher multiplicity the associated cluster algebra becomes infinite. Very recently, this obstacle has been overcome, at least for the case of $n=8$ particles, thanks to the intriguing interplay between cluster algebras and tropical Gra\ss mannians~\cite{Drummond:2019cxm,Arkani-Hamed:2019rds,Henke:2019hve}.

\acknowledgments
\noindent
GP and LD would like to thank Mark Spradlin and Thomas Harrington for collaboration in closely related topics. GP would also like to thank the organizers of CORFU2019 Workshop, in particular Jan Kalinowski and George Zoupanos, for the invitation and the stimulating atmosphere. The research presented in this contribution has received funding from the European Union's Horizon 2020 research and innovation programme under the Marie Sklodowska-Curie grant agreements
No. 764850 and No. 793151,  the National Science Foundation under Grant No.\ NSF PHY17-48958, the US Department of Energy under contract DE--AC02--76SF00515, the National Science and Engineering
Council of Canada, the Canada Research Chair program the
Fonds de Recherche du Qu\'ebec -- Nature et Technologies, Grants \mbox{(No.\ 757978)} and \mbox{(No.\ 648630} from the European Research Council, the Munich Institute for Astro- and Particle Physics (MIAPP) of the DFG cluster of excellence ``Origin and Structure of the Universe'',
the Danish National Research Foundation (DNRF91), a grant from the Villum Fonden, a grant from the Simons Foundation (341344, LA), a Carlsberg Postdoctoral Fellowship (CF18-0641), and a Humboldt Research Award (LD).

\appendix 
\section{Momentum twistors \& parametrizations}
Momentum (super) twistors may be expressed in terms of momenta and their supersymmetric partners, the supermomenta,\footnote{Where $\alpha,\dot \alpha=1,2$ are the usual $SL(2)$ spinor indices, and $A$ is an $R$-symmetry index.}
\be
p_i^{\alpha \dot\alpha} =
\lambda_i^\alpha \tilde{\lambda}_i^{\dot\alpha}
= x_{i+1}^{\alpha \dot\alpha} - x_{i}^{\alpha \dot\alpha}\,, \qquad
q_i^{\alpha A} = \lambda_i^{\alpha} \eta_i^A
= \theta_{i+1}^{\alpha A} - \theta_{i}^{\alpha A}\,,
\label{xthetadef}
\ee
by
\be\label{eq:supertwistors}
\mathcal{Z}_i = (Z_i \, | \, \chi_i)\,, \qquad
Z_i^{\alpha,\dot\alpha} =
(\lambda_i^\alpha , x_i^{\beta \dot\alpha}\lambda_{i\beta})\,,
\qquad
\chi_i^A= \theta_i^{\alpha A}\lambda_{i \alpha} \,.
\ee

We close by providing explicit momentum twistor parametrizations for the amplitudes considered in this contribution. These may be encoded in matrices where the matrix element $(ij)$ denotes the $i$-th component of the cyclically ordered momentum twistor $Z_j$.
\be
n=6:\quad\quad\quad 
\left(
\begin{array}{cccccc}
 -1 & 0 & 0 & 0 & 1 & 1+x_1 \\
 0 & 1 & 0 & 0 & 1+x_2 & 1+x_2+x_1 x_2 \\
 0 & 0 & -1 & 0 & 1+x_3+x_2 x_3 & 1+x_3+x_2 x_3+x_1 x_2 x_3 \\
 0 & 0 & 0 & 1 & 1 & 1 \\
\end{array}
\right)
\ee

For $n=7$, the first four momentum twistors are identical to the corresponding $n=6$ ones, whereas the rest are
\be
Z_5=\left(
\begin{array}{c}
 1 \\
 1+x_4+x_3 x_4 \\
 1+x_6+x_4 x_6+x_5 x_6+x_4 x_5 x_6+x_3 x_4 x_5 x_6 \\
 1 \\
\end{array}\right)\,,
\ee

\be
Z_6=\left(
\begin{array}{c}
 1+x_2 \\
 1+x_4+x_2 x_4+x_3 x_4+x_2 x_3 x_4 \\
 1+x_6+x_4 x_6+x_2 x_4 x_6+x_5 x_6+x_4 x_5 x_6+x_2 x_4 x_5 x_6+x_3 x_4 x_5 x_6+x_2 x_3 x_4 x_5 x_6 \\
 1 \\
\end{array}
\right)\,,
\ee
and

\be
Z_7=\left(
\begin{array}{c}
 1+x_2+x_1 x_2 \\
 1+x_4+x_2 x_4+x_3 x_4+x_2 x_3 x_4+x_1 x_2 x_3 x_4 \\
 1+x_6+x_4 x_6+x_2 x_4 x_6+x_5 x_6+x_4 x_5 x_6+x_2 x_4 x_5 x_6+x_3 x_4 x_5 x_6+x_2 x_3 x_4 x_5 x_6+x_1 x_2 x_3 x_4 x_5 x_6 \\
 1 \\
\end{array}
\right)
\ee

\bibliographystyle{JHEP}

\bibliography{Corfu_refs.bib}

\end{document}